\documentclass{IEEEtran} %%[conference]
\usepackage{amsmath}
\usepackage{amsfonts}
\usepackage{amssymb}
\usepackage{amsbsy}
\usepackage{graphicx}
\usepackage{times}
\usepackage{rotating}
\usepackage{bm}
\usepackage{color}
\usepackage{cite}
\usepackage{enumerate} 
\usepackage[letterpaper,top=.75in,bottom=0.75in,right=0.75in,left=0.75in]{geometry}

\newtheorem{definition}{Definition}
\newtheorem{theorem}{Theorem}
\newtheorem{pro}{Proposition}

\newtheorem{lemma}{Lemma}
\newtheorem{remark}{Remark}%[section]

\newcommand{\be}{\begin{equation}}
\newcommand{\ee}{\end{equation}}
\newcommand{\ist}{\hspace*{.3mm}}
\newcommand{\rmv}{\hspace*{-.3mm}}
\providecommand{\abs}[1]{\lvert#1\rvert}
\providecommand{\norm}[1]{\lVert#1\rVert}

\def\L{L}

\def\R{R}
\def\Q{Q}

\def\T{T}

\def\0v{\boldsymbol{0}}
\def\Iv{\boldsymbol{I}}

\def\sv{\boldsymbol{s}}

\def\uv{\boldsymbol{u}}
\def\uvt{\tilde{\uv}}
\def\vv{\boldsymbol{v}}

\def\xv{\boldsymbol{x}}

\def\yv{\boldsymbol{y}}
\def\yvb{\bar{\yv} }  %%%%%%%%%%%%%%%%%%

\def\xiv{\boldsymbol{\xi}}
\def\muv{\boldsymbol{\mu}}

\def\Am{\boldsymbol{A}}

\def\Bm{\boldsymbol{B}}
\def\Cm{\boldsymbol{C}}

\def\Jm{\boldsymbol{J}}
\def\Pm{\boldsymbol{P}}
\def\Mm{\boldsymbol{M}}
\def\Qm{\boldsymbol{Z}}

\def\Xm{\boldsymbol{X}}

\def\Xim{\boldsymbol{\Xi}}
\def\Ximt{\tilde{\Xim}}

\def\Sigm{\boldsymbol{\Sigma}}

\def\r0v{\boldsymbol{\mathsf{0}}}

\def\rhv{\boldsymbol{\mathsf{h}}}

\def\rnv{\boldsymbol{\mathsf{n}}}

\def\rsv{\boldsymbol{\mathsf{s}}}

\def\ruv{\boldsymbol{\mathsf{u}}}

\def\rvv{\boldsymbol{\mathsf{v}}}

\def\rxv{\boldsymbol{\mathsf{x}}}

\def\ryv{\boldsymbol{\mathsf{y}}}
\def\ryvb{\bar{\ryv} }  %%%%%%%%%%%%%%%%%%

\def\rAm{\boldsymbol{\mathsf{A}}}

\def\rXm{\boldsymbol{\mathsf{X}}}

\def\IN{\mathbb{N}}
\def\IC{\mathbb{C}}
\def\IR{\mathbb{R}}
\def\IZ{\mathbb{Z}}
\def\sA{\mathcal{A}}
\def\sB{\mathcal{B}}
\def\sC{\mathcal{C}}
\def\sD{\mathcal{D}}

\def\sF{\mathcal{F}}
\def\sI{\mathcal{I}}
\def\sG{\mathcal{G}}
\def\sJ{\mathcal{J}}
\def\sM{\mathcal{M}}

\def\sL{\mathcal{L}}

\def\sS{\mathcal{S}}

\def\sP{\mathcal{P}}
\def\sPt{\tilde{\sP}}
\def\sQ{\mathcal{Z}}
\def\sU{\mathcal{U}}
\def\sV{\mathcal{V}}
\def\sW{\mathcal{W}}
\providecommand{\absdet}[1]{\lvert#1\rvert}
\def\diag{\operatorname{diag}}
\def\rank{\operatorname{rank}}
\def\lcm{\operatorname{lcm}}
\def\trans{{\operatorname{T}}}
\def\E{\mathbb{E}}

\allowdisplaybreaks

\begin{document}

%% \IEEEoverridecommandlockouts

\title{\vspace{.25in}A Lower Bound on the Noncoherent Capacity Pre-log for the MIMO Channel with Temporally Correlated \vspace{.5mm}Fading}

\author{
\IEEEauthorblockN{G\"unther Koliander$^1\rmv$, Erwin Riegler$^1\rmv$, Giuseppe Durisi$^2\rmv$, Veniamin~I.~Morgenshtern$^3\rmv$, and Franz~Hlawatsch$^1$}\\[3.5mm]
\IEEEauthorblockA{\normalsize
$^1$Institute of Telecommunications, Vienna University of Technology, 1040 Vienna, Austria\\
$^2$Department of Signals and Systems, Chalmers University of Technology, 41296 Gothenburg, Sweden\\
$^3$Department of Statistics, Stanford University, CA 94305, USA
}
\thanks{This work was supported by the WWTF under grant ICT10-066 (NOWIRE).}
\vspace*{-.1mm}
}                                                                                               

\maketitle

\thispagestyle{empty}

\begin{abstract}
We derive a lower bound on the capacity pre-log of a temporally correlated Rayleigh block-fading multiple-input multiple-output (MIMO) channel with $\T$ transmit antennas and $\R$ receive antennas in the noncoherent setting (no \emph{a priori} channel knowledge at the transmitter and the receiver). In this  model, the fading process changes independently across blocks of length $\L$ and is temporally correlated within each block  for each transmit-receive antenna pair, with a given rank $\Q$ of the corresponding correlation matrix. Our result implies that for almost all choices of the coloring matrix that models the temporal correlation, the pre-log can be lower-bounded by $\T(1-1/\L)$ for $\T\leq(\L-1)/\Q$ provided that $\R$ is sufficiently large.
%and the correlation matrices are chosen in a generic way. 
The widely used constant block-fading model is equivalent to the temporally correlated block-fading model with $\Q=1$ for the special case when the temporal correlation for each transmit-receive antenna pair is the same, which is unlikely to be observed in practice.
For the constant block-fading model, the capacity pre-log is given by $\T(1-\T/\L)$, which is smaller than our lower bound for the case $\Q=1$. Thus, our result suggests that the assumptions underlying the constant block-fading model lead to a pessimistic result for the capacity 
pre-log.
%that the capacity pre-log increases in the presence of generic correlation matrices.
%For $\Q=1$, our result shows that the assumptions in the widely used constant block-fading model, where the capacity pre-log is known to be upper-bounded by $\T(1-\T/L)$ for $\T\leq \L/2$ transmit antennas, seem to be too pessimistic. This is due to the fact that the constant block-fading model is equivalent to the $Q=1$ correlated block-fading model for the special case when all correlation matrices are equal, which is not a generic choice.  
%Our result implies that the capacity pre-log of a rank-$\Q$ MIMO channel coincides with the capacity pre-log of a rank-$1$ MIMO channel provided that the block length $\L>\T\Q$  and $\R$ is sufficiently large. 
\vspace*{-3mm}
\end{abstract}

%%%%%%%%%%%%%%%%%%%%%%%%%%%%%%%
\section{Introduction} \label{sec:intro}   
%%%%%%%%%%%%%%%%%%%%%%%%%%%%%%%

\vspace*{.5mm}

We analyze the capacity of a Rayleigh block-fading mul\-ti\-ple-input multiple-output (MIMO) channel in the noncoherent setting where the transmitter and the receiver are aware of the channel statistics but have no \emph{a priori} channel state information.
In this setting, the penalty on 
capacity\footnote{In 
this paper, the term \emph{capacity} refers to capacity in the noncoherent setting.}
%% we will refer to capacity in the noncoherent setting simply as capacity.} 
incurred by allocating resources to channel estimation is automatically accounted for.
We model channel variations in time by
%% using 
the \emph{temporally correlated block-fading model} 
%% first 
introduced in~\cite{live04}.     
According to this  model, the fading process takes on independent realizations across blocks of length $\L$; however, for each transmit-receive antenna pair, it is correlated within each block with a given rank $\Q$ of the corresponding $\L\times\L$ correlation matrix. 

The capacity of the temporally correlated block-fading channel is not known even in the single-input single-output (SISO) case.
The capacity pre-log, which is defined as the ratio 
%% between 
of the capacity 
%% and 
to the logarithm of the signal-to-noise ratio (SNR) as the SNR goes to infinity, has been characterized in~\cite{live04} for the SISO case and in~\cite{modubo10,rimodulistbo11,yadumori11} for the single-input multiple-output (SIMO) case.
For \emph{regular stationary} fading processes, the capacity of the MIMO channel has been studied in~\cite{lamo03}. 
It was proved that, in this case, the capacity grows only doubly-logarithmically due to the regular fading assumption.
%
%For \emph{nonregular} stationary fading processes, a lower bound on the sum-rate capacity pre-log for the MIMO MAC has been derived in~\cite{asyhari11-07a}. However, it is not known whether this
%%% the resulting lower 
%bound is tight.
For \emph{nonregular stationary} fading processes, the MIMO capacity pre-log is not known to date.
%We note that the temporally correlated block-fading channel corresponds to a fading process that is nonstationary within each block but ``stationary between blocks.''

In this paper, we derive a lower bound on the capacity pre-log of a rank-$\Q$ temporally correlated block-fading MIMO channel with block length $\L$, $\T$ transmit antennas, and $\R$ receive antennas.
We show that the pre-log is lower-bounded by $\T (1-1/\L)$ for $\T\leq (\L-1)/\Q$ provided that $\R\geq \T(\L-1)/(\L-\T\Q)$. This lower bound can be achieved for almost all (a.a.)\ choices---i.e., up to a set of measure zero---of the coloring matrix that models the temporal correlation for the
%% all 
transmit-receive antenna pairs.

Our result is particularly surprising when compared to the capacity for the constant block-fading model as derived by Zheng and Tse~\cite{zhengtse02}. The constant block-fading model is a special case of the temporally correlated block-fading model for $\Q=1$ that is obtained when the correlation matrices for all transmit-receive antenna pairs are assumed to be equal, which is unlikely to be observed in practice.
%Their model corresponds to our system model for $\Q=1$ and a specific choice of correlation matrices. 
Zheng and Tse showed that the pre-log for the constant block-fading model is $M^* (1-M^*/\L)$ with $M^*\triangleq \min\{\T, \R, \lfloor\L/2\rfloor\}$, which is less than or equal to $\L/4$.  In the temporally correlated block-fading model for $\Q=1$, our lower bound on the pre-log is $\L-2+1/\L$ if $\T=\L-1$ and $\R=(\L-1)^2$ for a.a.\ coloring 
matrices.\footnote{Note %%%%%%%%%%%%
that the coloring matrix corresponding to
%% resulting in 
the constant block-fading model belongs to the set of measure zero where this bound does not hold.} %This implies that the correlation in the constant block-fading model does not belong to the generic correlations for which our bound holds and, hence, seems to be too pessimistic.
This shows that a much higher pre-log can be achieved and, hence, the results predicted by the constant block-fading model are pessimistic. 
%This shows that this more complex correlation can increase the pre-log.
 
%In the constant block fading model~\cite{zhengtse02}, which corresponds to the $\Q=1$ case with the same statistics for each transmit-receive antenna pair, the pre-log can attain at most $\L/4$. Our result gives a lower bound of $(\L-1)^2/\L>\L-2$ for the pre-log in the correlated setting for $\Q=1$. This shows that for a system where there is no restriction to the number of antennas, different correlation matrices increase the pre-log up to a factor of $4$.

Apart from our main result, the methods employed in its proof may be of independent interest. We use a generalized change-of-variables theorem for integrals in combination 
%\newgeometry{top=.75in,bottom=.75in,right=0.75in,left=.75in}
with B\'ezout's theorem \cite[Proposition B.2.7]{VdE00} to establish certain transformation properties of differential entropy under finite-to-one mappings. Furthermore, we use an important property of subharmonic functions to lower-bound the integral of a certain real analytic function. In the SIMO case, a similar problem was recently solved using an algebraic-geometry method  \cite{rimodulistbo11}. Our alternative method works in a more general setting and, thus, may also be useful for bounding differential entropy terms appearing in other problems.

The rest of this paper is organized as follows. 
The system model is presented in Section \ref{sec:syst}.
The lower bound on the capacity pre-log is stated and discussed in Section \ref{sec:bound}.
A proof of the lower bound is provided in Sections \ref{sec:proofprop} and  \ref{sec:boundh} and in three appendices.
%% Some ??? are provided in an 
\vspace{1mm}
%% Appendix.

%% \newpage %%%%%%%

%% \subsection{Notation}\label{notation}
\emph{Notation}:
Sets are denoted by calligraphic letters (e.g., $\sI$), and $|\mathcal{I}|$ 
denotes the cardinality of $\mathcal{I}$. Sets of sets are denoted by fraktur letters (e.g., $\mathfrak{M}$).
We use the notation $[M\!:\!N]
\triangleq \{M, M\!+\rmv 1,\dots,N\}$ for $M, N \!\in\! \IN$.
Boldface uppercase (lowercase) letters denote matrices (vectors). Sans serif letters denote random quantities,  e.g., $\rAm$ is a random matrix and $\rxv$ is a random vector. 
%All conventions for algebraic manipulations that are introduced for deterministic quantities will also be used for random quantities.
The superscripts ${}^{\operatorname{T}}$ and ${}^{\operatorname{H}}$ stand for transposition and 
Hermitian transposition, respectively. 
The all-zero matrix or vector of appropriate size is written as $\0v$, and the $M\times M$ identity matrix as
$\Iv_{M}$. For a matrix $\Am\in\IC^{M\times N}\rmv$, the element in the $i$th row and $j$th column 
is denoted by $a_{i,j}$. 
We denote by ${[\Am]}_{\sI}^{\sJ}$, where $\sI\subseteq [1\!:\!M]$ and $\sJ\subseteq [1\!:\!N]$,  
the $|\sI|\times|\sJ|$ submatrix of $\Am$ containing the elements $a_{i,j}$ with $i \!\in\! \sI$ and $j \!\in\! \sJ$;
furthermore,
${[\Am]}_{\sI} \!\triangleq {[\Am]}_{\sI}^{[1:N]}$ and ${[\Am]}^{\sJ} \! \triangleq {[\Am]}_{[1:M]}^{\sJ}$. 
We denote by ${[\xv]}_\sI\in\IC^{|\sI|}$ the subvector of $\xv$ containing the elements
$x_i$ with $i\in\sI$.
The diagonal matrix with the elements of 
$\xv$ in its main diagonal is denoted by $\diag(\xv)$.
We define
$\diag(\Am_1,\dots,\Am_K)$ as the block diagonal matrix with the matrices $\Am_1,\dots,\Am_K$ on the main block diagonal.
The modulus of the determinant of a square matrix $\Am$ is denoted by $\abs{\Am}$.
%The Kronecker product 
%is denoted by $\Am\otimes\Bm$ with the con\-vention that 
%$\Am\otimes\Bm\Cm \triangleq \Am\otimes (\Bm\Cm)$. 
For $x \!\in\!\IR$, $\lfloor x\rfloor\triangleq \max\{m \!\in\! \IZ \!\mid\! m \!\leq\! x\}$ and 
$\lceil x\rceil\triangleq \min\{m \!\in\! \IZ \!\mid\! m \!\geq\! x\}$. 
We write $\E[\cdot]$ for the expectation operator,
and $\mathcal{CN}(\muv,\Sigm)$ for the distribution of a jointly proper Gaussian random vector with mean $\muv$ and covariance matrix $\Sigm$.
The Jacobian matrix of a differentiable function $\phi$ is denoted by $\Jm_{\phi}$.

\vspace{.5mm}

%%%%%%%%%%%%%%%%%%%%%%%%%%%%%%%
\section{System Model} \label{sec:syst}   
%%%%%%%%%%%%%%%%%%%%%%%%%%%%%%%

\vspace{1mm}

We consider a MIMO channel with $\T$ transmit and $\R$ receive antennas.
%% \footnote{The generalization to multiple transmit antennas per user is straightforward and omitted to keep the notation simple.} 
%(The generalization to multiple transmit antennas per user is straightforward and omitted to keep the notation simple.)
The fading process associated with
%% between 
each transmit-receive antenna pair conforms to the 
%% same 
temporally correlated block-fading model \cite{live04}, 
which results in the following channel input-output relations within a given block of length~$\L$:
\be\label{eq:model1}
\ryv_{r} \ist=\, \sqrt{\frac{\rho}{\T}} \!\sum_{t\in [1:\T]} \! \diag(\rhv_{r,t}) \, \rxv_{t} \ist+\, \rnv_{r} \,,\quad r\in [1\!:\!\R] \,.
\vspace{-.4mm}
\ee
Here,
$\rxv_{t}\in\IC^{\L}$ is the signal vector transmitted by the $t$th transmit antenna; 
$\ryv_{r}\in\IC^{\L}$ is the vector received by the $r$th receive antenna; 
\pagebreak %%%%%%%%%
$\rhv_{r,t}\sim\mathcal{CN}(\0v,\Qm_{r,t}\Qm_{r,t}^{\operatorname{H}})$, where $\Qm_{r,t}\in\IC^{\L\times\Q}$ with $\Q\triangleq\rank(\Qm_{r,t}\Qm_{r,t}^{\operatorname{H}})$, 
is the vector of channel coefficients between the $t$th transmit antenna and the $r$th receive antenna;
$\rnv_{r}\sim\mathcal{CN}(\0v,\Iv_{\L})$ is the 
%% additive 
noise vector at the $r$th receive antenna; and
$\rho\in \IR^+$ is the SNR. 
The vectors $\rhv_{r,t}$ and $\rnv_{r}$ are assumed to be mutually independent and independent across 
$r\in [1\!:\!\R]$ and $t\in [1\!:\!\T]$, and to change in an independent fashion from block to block (``block-memoryless'' assumption). 
The transmitted signal vectors $\rxv_{t}$ are assumed to be independent of the vectors $\rhv_{r,t}$ and $\rnv_{r}$.
% and to satisfy the average per-user power constraints 
%\be
%\label{eq:powerconstr}
%\sum_{t\in \Tind_u} \rmv\E[\|\xv_{t}\|^2] \ist\leq\ist \L\abs{\Tind_u} \,, \quad u\in [1\!:\!\U]\,, 
%\ee
%where $\Tind_u$ denotes the index set of the transmit antennas belonging to user $u\in [1\!:\!\U]$. 
We note that the channel coefficient vectors can be written 
%% in whitened form 
as 
\vspace*{-.5mm}
\[
\rhv_{r,t} \ist=\ist \Qm_{r,t}\rsv_{r,t}\,,
\]
 with the $\Q$-dimensional whitened vectors $\rsv_{r,t}\sim\mathcal{CN}(\0v,\Iv_{Q})$.

Setting
$\ryv\!\triangleq\rmv(\ryv_{1}^{\operatorname{T}},\dots,\ryv_{\R}^{\operatorname{T}})^{\operatorname{T}}\rmv\in\IC^{\R\L}\!$ and
$\rnv\!\triangleq\rmv(\rnv_{1}^{\operatorname{T}},\dots,\rnv_{\R}^{\operatorname{T}})^{\operatorname{T}}$\linebreak %%%%%
$\rmv\in\IC^{\R\L}$, 
the $\R$ input-output relations \eqref{eq:model1} can be written more compactly 
%\pagebreak %%%%%%
as
%\vspace*{-1.5mm}
\be\label{eq:model2}
\ryv\,=\ist \sqrt{\frac{\rho}{\T}} \ist \ryvb \ist+\ist \rnv \,,
\vspace*{-1.5mm}
\ee
with
\vspace*{2mm}
\be\label{eq:ybar}
\ryvb \ist\triangleq\! \sum_{t\in[1:\T]}
\!\begin{pmatrix}
\rXm_t\Qm_{1,t} \hspace*{-2mm} &\hspace*{-4mm}&\hspace*{-2mm} \\[-1.5mm]
\hspace*{-2mm} & \hspace*{-2mm}\ddots\hspace*{-2mm} \\[-.3mm]
\hspace*{-2mm} & \hspace*{-4mm} & \hspace*{-2mm} \rXm_t\Qm_{\R,t}
\end{pmatrix}
\rmv \rsv_t \rmv\in\IC^{\R\L} \ist,
%\vspace{-.7mm}
\ee
where we have defined 
$\rXm_{t}\triangleq\diag(\rxv_{t})\in\IC^{\L\times\L}$
and $\rsv_{t}\triangleq(\rsv_{1,t}^{\operatorname{T}},\dots,$ $\rsv_{\R,t}^{\operatorname{T}})^{\operatorname{T}}\rmv\in\IC^{\R\Q}$.
For later use, we also define 
$\rxv\triangleq(\rxv_{1}^{\operatorname{T}},\dots,\rxv_{\T}^{\operatorname{T}})^{\operatorname{T}}\rmv\in\IC^{\T\L}$, 
$\rsv\triangleq(\rsv_{1}^{\operatorname{T}},\dots,\rsv_{\T}^{\operatorname{T}})^{\operatorname{T}}\rmv\in\IC^{\T\R\Q}$, and
\[
\Qm \ist\triangleq\ist \begin{pmatrix}
\Qm_{1,1} & \hspace*{-2mm}\cdots\hspace*{-2mm} & \Qm_{1,\T}\\[-.8mm]
\vdots & & \vdots \\
\Qm_{\R,1} & \hspace*{-2mm}\cdots\hspace*{-2mm} & \Qm_{\R,\T}
\end{pmatrix}
\rmv\in \IC^{\R\L\times \T\Q}.
\]
We will refer to $\Qm$ as the \emph{coloring matrix}.

%% \vspace{.5mm}

%%%%%%%%%%%%%%%%%%%%%%%%%%%%%%%
\section{A Lower Bound on the Capacity Pre-log} \label{sec:bound}   
%%%%%%%%%%%%%%%%%%%%%%%%%%%%%%%

\vspace{1mm}

Because of the block-memoryless assumption, the coding theorem in \cite[Section 7.3]{Gallager68} implies that the capacity of the channel \eqref{eq:model2} is given by 
\be\label{eq:capacity}
C(\rho) \ist=\ist \frac{1}{\L}\sup_{f}  I(\rxv \ist;\ryv) \,.
\ee
Here, $I(\rxv \ist;\ryv)$ denotes mutual information \cite[p.\,251]{Cover91} and the supremum is taken over all input distributions $f$ on $\IC^{\T\L}$
that satisfy the average power constraint
\[
%\label{eq:powerconstr}
\E[\|\rxv\|^2] \ist\leq\ist \T\L \,.
\]
The capacity pre-log is then defined as  
\be\label{eq:prelog1}
\chi \ist \triangleq\lim_{\rho\to\infty}\frac{C(\rho)}{\log(\rho)} \,.
\vspace{.5mm}
\ee

We will obtain the main result of this paper, which is stated in Theorem \ref{th:maintheorem} below, by maximizing with respect to $\T$ the lower bound on $\chi$ 
given in the following \vspace{1.5mm}
proposition.

\begin{pro}\label{pro:prelim}
For $\T \!\leq\!\R$, there exists a set $\sQ\!\subseteq\! \IC^{\R\L\times\T\Q}$ with a complement of Lebesgue measure zero such that for each coloring matrix $\Qm\in \sQ$, 
the capacity pre-log of the chan\-nel \eqref{eq:model1} 
satisfies 
\pagebreak %%%%%%
\vspace*{-4.5mm}
\be\label{eq:prelim}
\chi \,\geq\, \chi_{\operatorname{low}}(\T) \,\triangleq\, \min\rmv\bigg\{\T\bigg(1 \rmv-\rmv \frac{1}{\L}\bigg), \R\bigg(1 \rmv-\rmv \frac{\T\Q}{\L}\bigg)\bigg\} \,.
\ee
\end{pro}
\begin{IEEEproof}[\hspace{-1em}Proof]
See Section \ref{sec:proofprop}.
\end{IEEEproof}

The main result of this paper is stated in the following 
\vspace{1.5mm}
theorem. 

\begin{theorem}\label{th:maintheorem}
There exists a set $\sQ \!\subseteq\! \IC^{\R\L\times\T\Q}\rmv$ with a com\-ple\-ment of Lebesgue measure zero such that for each coloring\linebreak %%%%% 
matrix $\Qm \!\in\! \sQ$, the capacity pre-log of the channel \eqref{eq:model1} 
\vspace{-.1mm}
satisfies 
\be\label{eq:maintheorem}
\chi \,\geq\, \chi^*_{\operatorname{low}} \,\triangleq\,
\begin{cases}
\T\bigg(1 \rmv-\rmv \dfrac{1}{\L}\bigg) & \text{ if } \T\leq\T_{\operatorname{opt}} \\
\eta & \text{ if } \T>\T_{\operatorname{opt}}\,,
\end{cases}
\vspace{-2mm}
\ee
where
\vspace{.5mm}
\be\label{eq:eta}
\eta \,\triangleq\, \max\bigg\{ \R\bigg(1-\frac{\lceil \T_{\operatorname{opt}}\rceil\Q}{\L}\bigg), \lfloor\T_{\operatorname{opt}}\rfloor\bigg(1\rmv-\rmv \frac{1}{\L}\bigg) \bigg\}
\vspace{-.2mm}
\ee
and
\vspace{1.5mm}
\be\label{eq:topt}
\T_{\operatorname{opt}} \,\triangleq\, \frac{\R\L}{\L+\R\Q-1} \,\leq\, \min\{\R, \L/\Q\} \,.
\vspace{1mm}
\ee
\end{theorem} 

\begin{IEEEproof}[\hspace{-1em}Proof]
%We can obtain a lower bound on the pre-log for $\T$ transmit antennas by reducing the number of transmit antennas (this can always be achieved by switching off some of them).
We obtain a lower bound on the pre-log for $\T$ transmit antennas by maximizing $\chi_{\operatorname{low}}$ 
with respect to the number of effectively used transmit antennas (note that we can always switch off some antennas).
Thus, we will take the maximum of $\chi_{\operatorname{low}}(\T^*)$ in \eqref{eq:prelim} with respect to $\T^*\leq \min\{\T,\R\}$. Here, $\T^*$ is also restricted by $\T^*\leq\R$ because Proposition \ref{pro:prelim} holds only for $\T^*\leq \R$.
Because $\chi_{\operatorname{low}}(\T^*)$ is the minimum of two quantities where the first is monotonically increasing in $\T^*$ 
and the second is monotonically decreasing in $\T^*\rmv$, it attains its maximum at the intersection point $\T_{\operatorname{opt}}$ defined in \eqref{eq:topt}.
If $\T\leq\T_{\operatorname{opt}}$, we have 
$
\chi_{\operatorname{low}}(\T) = \T(1-1/\L)
$,
which proves the first case in \eqref{eq:maintheorem}.
For $\T>\T_{\operatorname{opt}}$, we have to take into account that $\T_{\operatorname{opt}}$ might not be a natural number. Thus, we have to take the maximum of $\chi_{\operatorname{low}}(\lfloor\T_{\operatorname{opt}}\rfloor)$ and $\chi_{\operatorname{low}}(\lceil \T_{\operatorname{opt}}\rceil)$, which turns out to be $\eta$ in \eqref{eq:eta}. This shows the second case in \eqref{eq:maintheorem} and concludes the proof.
\end{IEEEproof}

\begin{remark}
The set $\sQ$ will be specified in Definition~\ref{def:qind} in Section~\ref{sec:boundh}. 
\end{remark}

\begin{remark}
For a fixed $\R$, the maximum value of $\chi^*_{\operatorname{low}}$ in \eqref{eq:maintheorem} is obtained by using either $\lfloor \T_{\operatorname{opt}}\rfloor$ or $\lceil \T_{\operatorname{opt}}\rceil$ transmit antennas. This implies that the optimal number of transmit antennas is 
%% always 
upper-bounded by $\lceil \T_{\operatorname{opt}}\rceil$.
\end{remark}

\begin{remark}
For $\L \rmv=\rmv \Q$, $\chi^*_{\operatorname{low}}$ is equal to zero and hence trivial.
\end{remark}

\begin{remark}
The lower bound $\chi^*_{\operatorname{low}}$ in \eqref{eq:maintheorem} can be expressed as
\[
\chi^*_{\operatorname{low}} \ist=\, \min\bigg\{\T\bigg(1 \rmv-\rmv \frac{1}{\L}\bigg), \eta\bigg\}\,.
\]
\end{remark}

\begin{remark}
$\chi^*_{\operatorname{low}}$ can be at most $\lfloor(\L \!-\! 1)/\Q\rfloor(1 \rmv-\rmv 1/\L)$. 
This value of $\chi^*_{\operatorname{low}}$ is attained for $\T=\lfloor(\L \!-\! 1)/\Q\rfloor$ and $\R=\lceil(\L \!-\! 1)^2/\Q\rceil$.
%Since $\T_{\operatorname{opt}}<\L/\Q$ for any $\R$ and since for $\lceil\T_{\operatorname{opt}}\rceil\geq \L/\Q$ the quantity $\eta$ in \eqref{eq:eta} reduces to $\lfloor\T_{\operatorname{opt}}\rfloor(1-1/\L)$, $\chi^*_{\operatorname{low}}$ can be at most $\lfloor(\L-1)/\Q\rfloor(1-1/\L)$. This value of $\chi^*_{\operatorname{low}}$ is attained for $\T=\lfloor(\L-1)/\Q\rfloor$ and $\R=\lceil(\L-1)^2/\Q\rceil$.
\end{remark}

\begin{remark}
By \eqref{eq:topt}, the condition $\T \!\leq\! \T_{\operatorname{opt}}$  in \eqref{eq:maintheorem} is equivalent to $\R\geq \T(\L \!-\! 1)/(\L \rmv-\rmv \T\Q)$. 
Thus, for a fixed $\T \!<\rmv \L/\Q$, we can always obtain $\chi^*_{\operatorname{low}} \!= \T(1 \rmv-\rmv 1/\L)$ by using a sufficiently large $\R$.
\end{remark}

\begin{remark}
%Note that for $\Q=1$ our system model does not reduce to the constant block fading model . 
If all matrices $\Qm_{r,t}$ for $r\in[1\!:\!\R]$ and  $t\in [1\!:\!\T]$ coincide, the temporally correlated block-fading model for $\Q=1$ reduces to the constant block-fading model studied in~\cite{zhengtse02}. The pre-log in the constant block-fading model is $M^* (1-M^*/\L)$, where $M^*\triangleq \min\{\T, \R, \lfloor\L/2\rfloor\}$; therefore, it is upper-bounded by $\L/4$. On the other hand, Theorem \ref{th:maintheorem} for $\Q=1$ implies that the pre-log for
%% in 
the correlated block-fading model is lower-bounded by (cf. \eqref{eq:prelim})
%% \vspace{-.5mm}
\be\label{eq:boundremark}
\chi\geq \min\bigg\{\T\bigg(1-\frac{1}{\L}\bigg), \R\bigg(1-\frac{\T}{\L}\bigg)\bigg\}\,,
%% \vspace{-.5mm}
\ee
for a.a.\ coloring matrices $\Qm \!\in\! \sQ$.
In particular, for $\T \rmv=\rmv \L \rmv-\! 1$ and $\R \rmv=\rmv (\L \!-\! 1)^2$, the lower bound in \eqref{eq:boundremark} becomes $\L-2+1/\L$. 
Thus, for a.a.\ choices of coloring matrices, the pre-log is much higher than in the constant block-fading 
model.\footnote{This %%%%%%%%%%
implies that the coloring matrices corresponding
%% resulting in system models equivalent 
to the constant block-fading model belong to the complement of 
$\sQ$ (which has Lebesgue measure zero and is unlikely to be observed in practice).} %%%%%%%%%% 
Hence, the results predicted by the constant block-fading model are pessimistic. 
%This implies that that $\Qm$ consisting only of ones does not belong to $\sQ$ and, hence, the constant block-fading model seems to be too pessimistic.
\end{remark}

%\begin{remark}\label{re:mimomac}
%Due to the fact that the sum-rate capacity of a MIMO MAC channel with a total number of transmit antennas equal to the number of transmitters in the SIMO MAC is calculated as in \eqref{eq:capacity} but with the supremum taken over joint distributions for all antennas of each user, it is always larger than or equal to the sum-rate capacity in the SIMO MAC case.
%Therefore, the lower bound also holds for a MIMO MAC setting.
%\end{remark}
%The highest lower bound attained by Theorem \ref{th:maintheorem} for given $\L$ and $\Q$ is described in the following corollary.
%\begin{cor}
%The capacity pre-log \eqref{eq:prelog1} of the channel \eqref{eq:model1} with $\T=\lceil\L/\Q\rceil-1$ transmit antennas, $\R = (\L-1)^2/\Q$ receive antennas, and $\Qm\in \sQ$ satisfies 
%\be
%\chi\geq \bigg(\bigg\lceil\frac{\L}{\Q}\bigg\rceil-1\bigg)\bigg(1-\frac{1}{\L}\bigg)\,.
%\ee
%\end{cor}

\vspace{-1mm}

%%%%%%%%%%%%%%%%%%%%%%%%%%%%%%%
\section{Proof of Proposition \ref{pro:prelim}} \label{sec:proofprop}
%%%%%%%%%%%%%%%%%%%%%%%%%%%%%%%

\vspace{.5mm}

For $\L\leq\T\Q$, the inequality in \eqref{eq:prelim} is trivially true, because in this case $\chi_{\operatorname{low}}\leq 0$. 
Therefore, it remains to prove \eqref{eq:prelim} for $\L>\T\Q$, which will thus be assumed in the following.
By \eqref{eq:capacity}, the capacity can be lower-bounded as  
$C(\rho) \ist\geq\ist (1/\L) I(\rxv \ist;\ryv)$
with the specific input distribution $\rxv \sim \mathcal{CN}(\0v,\Iv_{\T\L})$. Inserting this lower bound into \eqref{eq:prelog1} then gives
\be \label{eq:preloggauss}
\chi \, \geq\ist \frac{1}{\L} \lim_{\rho\to\infty}\frac{I(\rxv \ist;\ryv)}{\log(\rho)} \,.
\ee
%This implies that \cite[Lemma 6.7]{lamo03}
%\[
%% \label{eq:propx}
%\E\ist[\ist \log(| x_{t,i} |)] \ist>\ist -\infty \,,\quad t\in [1\!:\!\T] \, ,\; i\in[1\!:\!\L] \,, 
%\]
%where $x_{t,i}$ denotes the $i$th element of $\rxv_{t}$.
In what follows, we thus assume that $\rxv \sim \mathcal{CN}(\0v,\Iv_{\T\L})$.

We have $I(\rxv \ist;\ryv)=h(\ryv)-h(\ryv\ist|\ist\rxv)$ with $h$ denoting differential entropy. Hence, we can lower-bound $I(\rxv \ist;\ryv)$ by 
%% first finding an 
upper-bounding $h(\ryv\ist |\ist\rxv)$ and lower-bounding $h(\ryv)$. Similar to \cite[Eq.~(8)]{modubo10}, we have
\be
\label{eq:boundhyx}
h(\ryv\ist |\ist \rxv) \,\leq\, \T\Q\R \ist \log(\rho) \ist+\ist \mathcal{O}(1) \,,
\ee
where ``$+\,\ist\mathcal{O}(1)$'' means ``up to a function of $\rho$ that is bounded for $\rho\to\infty$.''
%% is taken with respect to $\rho$, and 
Furthermore, similar to \cite[Eq.~(12)]{modubo10}, we have
\be\label{eq:boundhy}
%% \begin{split}
h(\ryv)
%&=h(\Pm\ryv,\Qm\ryv)\nonumber\\
%&=h(\Pm\ryv)+h(\Qm\ryv\mid\Pm\ryv)\nonumber\\
%&\geq h(\sqrt{\rho}\Pm\uv+\Pm\nv\mid\Pm\nv)+h(\Qm\ryv\mid\rsv,\rxv,\Pm\ryv)\nonumber\\
\ist\geq \Bigg(\rmv \sum_{r\in [1:\R]} \!\rmv |\sI_r|
%% \alpha_t
\Bigg) \log(\rho)
 + h (\Pm\ryvb )+c \,,
\ee
where $\ryvb$ was 
\vspace{.7mm}
defined in \eqref{eq:ybar},
\be \label{eq:Pm}
\Pm \ist\triangleq\, \diag \rmv\big( {[\Iv_{\L}]}_{\sI_1}  ,\dots,{[\Iv_{\L}]}_{\sI_{\R}} \big) \in \IC^{\sum_{r\in[1:\R]} \!|\sI_{r}| \times \R\L}\ist,
%% \label{eq:Pm}
%\Qm&\triangleq \diag({(\Iv_{\L})}_{[1\!:\!\L]\setminus\I_{1}},\dots,{(\Iv_{\L})}_{[1\!:\!\L]\setminus\I_{\R}}).
\vspace{.7mm}
\ee
the
%% for sets 
$\sI_r\subseteq [1\!:\!\L]$ for $r\in[1\!:\!\R]$ are certain subsets that will be specified later, and
$c$ is a finite constant. Note that in \eqref{eq:boundhy}, $h (\Pm\ryvb)$ and 
 $c$ do not depend on $\rho$. Using \eqref{eq:boundhyx} and \eqref{eq:boundhy} in $I(\rxv \ist;\ryv)=h(\ryv)-h(\ryv\ist|\ist\rxv)$, we obtain
\begin{align}
&I(\rxv \ist;\ryv) \ist\geq\ist \Bigg( \sum_{r\in [1:\R]} \!\rmv |\sI_r|-\T\Q\R \Bigg)\log(\rho)+ h (\Pm\ryvb)  \ist+\ist \mathcal{O}(1)\,.\nonumber\\[-4mm] 
\label{EQmutual}\\[-6.5mm]
\nonumber
\end{align}
The proposed lower bound on the pre-log in \eqref{eq:prelim} is established by inserting \eqref{EQmutual} into \eqref{eq:preloggauss} 
%\be\label{eq:preloggauss2}
%\chi \geq \frac{1}{\L}\Bigg(\T\L-\! \sum_{t\in [1:\T]}\! |\sP_t| + \lim_{\rho\to\infty}\frac{h \big(\Pm\ryvb\ist | \ist {[\rxv_1]}_{\sP_1}, \dots, {[\rxv_{\T}]}_{\sP_{\T}} \big)}{\log(\rho)}  \Bigg).
%\ee
%With
and choosing 
\pagebreak %%%%%%%
the sets $\{\sI_r\}_{r\in[1:\R]}$ such that
\be
\label{EQalpha}
\sum_{r\in [1:\R]} \!\rmv |\sI_r|
\,=\, \min\ist\{ \T\L-\T+\T\Q\R, \R\L\} \,,
%% \vspace{-.7mm}
\ee
provided that $h (\Pm\ryvb) \rmv >\rmv -\infty$.
It remains to show that there exist sets $\{ \sI_r \}_{r\in [1:\R]}$ satisfying \eqref{EQalpha} and a set $\sQ\subseteq\IC^{\R\L\times\T\Q}$ with a complement of Lebesgue measure zero for which $h (\Pm\ryvb) \rmv >\rmv -\infty$ for each $\Qm\in \sQ$. 
This will be done in the next section.

%% \newpage %%%%%%%%

\vspace{-1mm}

%%%%%%%%%%%%%%%%%%%%%%%%%%%%%%%
%%%%%%%%%%%%%%%%%%%%%%%%%%%%%%%
\section{Proof that $\,h (\Pm\ryvb) \rmv >\rmv -\infty$} \label{sec:boundh}   
%%%%%%%%%%%%%%%%%%%%%%%%%%%%%%%

\vspace{1mm}

Let us split the vector $\rxv$ into the vectors $\rxv_{\sP}\triangleq \big({[\rxv_1]}_{\sP_1}^{\operatorname{T}}, \dots,$\linebreak$ {[\rxv_{\T}]}_{\sP_{\T}}^{\operatorname{T}}\big)^{\operatorname{T}}\rmv$ and $\rxv_{\sD}\triangleq \big({[\rxv_1]}_{\sD_1}^{\operatorname{T}},\dots,{[\rxv_{\T}]}_{\sD_{\T}}^{\operatorname{T}}\big)^{\operatorname{T}}\rmv$, 
where $\sP_t\subseteq[1\!:\!\L]$ and $\sD_t\triangleq[1\!:\!\L]\backslash\sP_t$ for $t\in [1\!:\!\T]$.
Because $h (\Pm\ryvb\ist | \ist\rxv_{\sP})\leq h (\Pm\ryvb)$, it is sufficient to show that $h (\Pm\ryvb\ist | \ist\rxv_{\sP})>-\infty$.
As in \cite{rimodulistbo11}, we wish to relate 
%% the differential entropy 
$h (\Pm\ryvb\ist | \ist\rxv_{\sP})$ to the simpler quantity 
$h(\rsv,\rxv_{\sD}) \,=\, h(\rsv) \ist+ h(\rxv_{\sD})$. 
%Since the dimensions of $\Pm\ryvb$ and $(\rsv,\rxv)$ do not coincide 
%We now can relate the differential entropies $h (\Pm\ryvb\ist | \ist\rxv_{\sP})$ and $h(\rsv,\rxv_{\sD})$. 
This will be done via the family of $\rxv_{\sP}$-parametrized mappings
\be\label{eq:phi}
\phi_{\xv_{\sP}}\colon (\sv,\xv_{\sD}) \mapsto\, 
%\Pm \!
%% \Bigg(
%\sum_{t\in [1:\T]} \!(\Iv_{\R}\otimes\rXm_{t}\Qm) \ist\rsv_{t}
%% \Bigg)
%\ist=\ist 
\Pm\yvb \,,
\ee
where $\yvb$ is defined in \eqref{eq:ybar}, i.e.,
\be\label{eq:ybardisc}
\yvb \,= \sum_{t\in[1:\T]} \! \Xim_t\sv_t\,,
\vspace{-2mm}
\ee
with
\vspace{2mm}
\be\label{eq:defXi}
\Xim_t \ist\triangleq\ist
\begin{pmatrix}
\Xm_t\Qm_{1,t} \hspace*{-2mm} &\hspace*{-4mm}&\hspace*{-2mm} \\[-1.5mm]
\hspace*{-2mm} & \hspace*{-2mm}\ddots\hspace*{-2mm} \\[-.3mm]
\hspace*{-2mm} & \hspace*{-4mm} & \hspace*{-2mm} \Xm_t\Qm_{\R,t}
\end{pmatrix} \rmv\in \IC^{\R\L\times\R\Q} \ist.
\ee
According to \eqref{eq:ybardisc} and \eqref{eq:defXi}, the components of each vector-valued mapping $\phi_{\xv_{\sP}}$ are multivariate polynomials of degree 2.
The Jacobian matrix $\Jm_{\phi_{\xv_{\sP}}}$ of each mapping $\phi_{\xv_{\sP}}\rmv$
%\[
%\Jm_{\phi_{\xv_{\sP}}}(\sv,\xv_{\sD}) \ist\triangleq\ist \frac{\partial \Pm\yvb}{\partial (\sv,\xv_{\sD}) } \,,
%\in \IC^{\sum_{r\in[1:\R]} \!|\I_{r}| \times\T\ist (\Q\R\rmv+\rmv\L) - \sum_{t\in [1:\T]}\! |\sP_t|} 
%\]
is equal to
\begin{align}
\Jm_{\phi_{\xv_{\sP}}}\!(\sv,\xv_{\sD}) \ist & \ist=\ist \Pm \rmv \left( 
\Xim_1\,\cdots\, 
\Xim_{\T} \;
\begin{matrix}
\Am_{1,1} & \hspace*{-2.5mm}\cdots\hspace*{-2.5mm}  & \Am_{1,\T}  \\[-.8mm]
 \vdots    &   \hspace*{-2.5mm}\hspace*{-2.5mm}     & \vdots     \\ 
\Am_{\R,1} & \hspace*{-2.5mm}\cdots\hspace*{-2.5mm} & \Am_{\R,\T}
\end{matrix}
\right)\notag\\[.5mm]
& \quad \quad\quad\in \IC^{\sum_{r\in[1:\R]}\abs{\sI_r} \times \big(\T\Q\R \,+\ist \sum_{t\in[1:\T]}\abs{\sD_t}\big)},
\label{eq:Jacobian1}\\[-10mm]
\nonumber
\end{align}
where
\vspace*{.5mm}
\begin{align}
%gin{multline}\label{EQArt}
&\Am_{r,t} \ist\triangleq\ist
\big[\diag \big( a_{r,t}^{(1)} ,\dots,a_{r,t}^{(\L)} \big)
\big]^{\sD_t}\rmv, \quad\!\!
t\in[1\!:\!\T] \ist, \,r\in [1\!:\!\R]\,, \nonumber\\[1mm]
&\qquad\qquad \text{with} \;\; a_{r,t}^{(\ell)} \triangleq\ist [\Qm_{r,t}]_{\{\ell\}}\sv_{r,t} \,, \quad\!\! \ell\in [1\!:\!\L]\,.
\label{EQArt}\\[-4.5mm]
\nonumber
\end{align}
%nd{multline}
%% with $a_{r,t}^{(\ell)}\triangleq[\Qm_{r,t}]_{\{\ell\}}\sv_{r,t}$ for $\ell\in [1\!:\!\L]$.
Note that by \eqref{eq:Pm}, $\Jm_{\phi_{\xv_{\sP}}}\!(\sv,\xv_{\sD})$ can be 
%% equivalently 
written as 
\be
\hspace*{-2mm}\Jm_{\phi_{\xv_{\sP}}}\!(\sv,\xv_{\sD}) \,=\ist \left( 
\Ximt_1\,\cdots\, 
\Ximt_{\T} \;
\begin{matrix}
{[\Am_{1,1}]}_{\sI_1} & \hspace*{-2.5mm}\cdots\hspace*{-2.5mm}  & {[\Am_{1,\T}]}_{\sI_1}  \\[-.8mm]
 \vdots    &   \hspace*{-2.5mm}\hspace*{-2.5mm}     & \vdots     \\
{[\Am_{\R,1}]}_{\sI_{\R}} & \hspace*{-2.5mm}\cdots\hspace*{-2.5mm} & {[\Am_{\R,\T}]}_{\sI_{\R}}
\end{matrix}
\right) \rmv,
\label{eq:Jacobian3}
\vspace{-2mm}
\ee
where
\vspace{1.5mm}
\[
%% \label{eq:Jacobian3a}
\Ximt_t \ist\triangleq\ist
\begin{pmatrix}
{[\Xm_t\Qm_{1,t}]}_{\sI_1} \hspace*{-2mm} &\hspace*{-4mm}&\hspace*{-2mm} \\[-.7mm]
\hspace*{-2mm} & \hspace*{-2mm}\ddots\hspace*{-2mm} \\[0mm]
\hspace*{-2mm} & \hspace*{-4mm} & \hspace*{-2mm} {[\Xm_t\Qm_{\R,t}]}_{\sI_{\R}}
\end{pmatrix}.
\vspace*{1mm}
\pagebreak %%%%%%
\]
%The number of rows of $\Jm_{\phi_{\xv_{\sP}}}(\sv,\xv_{\sD})$ is 
%\be\label{eq:nrrows}
%\#\text{rows} =\!\!\sum_{r\in[1:\R]}\abs{\sI_r}\,,
%\ee
%and the number of columns is
%\be\label{eq:nrcolumns}
%\#\text{columns}=\T\Q\R+\T\L-\!\!\sum_{t\in[1:\T]}\abs{\sP_t}\,.
%\ee

Based on the family of mappings $\phi_{\xv_{\sP}}$ in \eqref{eq:phi}, the relation between $h (\Pm\ryvb\ist | \ist\rxv_{\sP})$ and $h(\rsv,\rxv_{\sD})$ can 
be established by using the definition of conditional differential entropy 
\cite[Chapter 8]{Cover91} and by applying the change-of-variables theorem for integrals under finite-to-one mappings\footnote{For a finite-to-one mapping, the inverse image of  each point in the codomain is a set  of finite cardinality.} \cite[Theorem~3.2.5]{fed69}. For this, we need to show that the family of mappings $\phi_{\xv_{\sP}}$
%% in \eqref{eq:phi}
is finite-to-one almost everywhere (a.e.)\ for a.a.\ choices of $\xv_{\sP}$.
We now define the set $\sQ$ for which this proof
%% derivation 
\vspace{1.5mm}
works.

\begin{definition}\label{def:qind}
Let $\sQ\subseteq\IC^{\R\L\times\T\Q}$ be the set of matrices $\Qm$ such that the following holds: 
 There exist a choice of sets $\{ \sI_r \}_{r\in [1:\R]}$ satisfying \eqref{EQalpha}, i.e.,
\be\label{eq:indi}
\sum_{r\in[1:\R]}\abs{\sI_r} \,=\, \min\{ \T\L-\T+\T\Q\R, \R\L\}\,,
\vspace{-.5mm}
\ee
and a choice of sets $\{\sP_t\}_{t\in [1:\T]}$ satisfying
\be\label{eq:indp}
\sum_{t\in[1:\T]}\abs{\sP_t} \,=\,  \max\{\T, \T\Q\R-(\R \rmv-\rmv \T)\L\}\,,
\vspace{-.5mm}
\ee
such that $\Jm_{\phi_{\xv_{\sP}}}\!(\sv,\xv_{\sD})$ is nonsingular a.e.\ for a.a.\ choices of \vspace{1.5mm}$\xv_{\sP}$.
\end{definition}

We will show presently that the set $\sQ$ is nonempty. In fact, it covers a.a.\ of $\IC^{\R\L\times\T\Q}$.

Condition \eqref{eq:indi} on $\{\abs{\sI_r}\}_{r\in[1:\R]}$ and condition \eqref{eq:indp} on $\{\abs{\sP_t}\}_{t\in[1:\T]}$ guarantee 
that the matrix $\Jm_{\phi_{\xv_{\sP}}}\!(\sv,\xv_{\sD})$ is square. More specifically, we have with %\eqref{eq:nrrows} and 
\eqref{eq:Jacobian1} that
\be\label{eq:nrrows}
\#\text{rows} \,= \!\sum_{r\in[1:\R]}\abs{\sI_r} \,=\, \min\{ \T\L-\T+\T\Q\R, \R\L\}\,,
\vspace{-.5mm}
\ee
where \eqref{eq:indi} was used, and 
\begin{align}
\#\text{columns} & \,=\, \T\Q\R+\!\!\sum_{t\in[1:\T]}\abs{\sD_t} \label{eq:nrcolumns_0} \\[.5mm]
& \,=\, \T\Q\R+\T\L-\!\!\sum_{t\in[1:\T]}\abs{\sP_t} \notag \\[.5mm]
& \,=\, \T\Q\R+\T\L - \max\{\T, \T\Q\R-(\R \rmv-\rmv \T)\L\} \notag \\[1mm]
%& = \min\{\T\Q\R+\T\L-\T, \T\Q\R+\T\L - \T\Q\R + (\R-\T)\L\} \notag \\
& \,=\, \min\{\T\Q\R+\T\L-\T, \R\L\}\,, \label{eq:nrcolumns}
\end{align}
where \eqref{eq:indp} was used. Thus, comparing \eqref{eq:nrrows} and \eqref{eq:nrcolumns}, we have
$\#\text{rows} = \#\text{columns}$. 
%Furthermore, comparing \eqref{eq:nrrows} and \eqref{eq:nrcolumns_0}, we also have
%\be\label{eq:columnsrows}
%% \#\text{rows}= \!\!\sum_{r\in[1:\R]}\abs{\sI_r} = \T\Q\R+\!\!\sum_{t\in[1:\T]}\abs{\sD_t} = \#\text{columns}\,.
%\sum_{r\in[1:\R]}\abs{\sI_r} \,=\, \T\Q\R \ist+\!\rmv\sum_{t\in[1:\T]}\abs{\sD_t} \,.
%\ee

The next lemma states that $\sQ$ satisfies one of the claims made in 
\vspace{1.5mm}
Proposition~\ref{pro:prelim}.

\begin{lemma}\label{LEMnotvanish}
The complement of the set $\sQ$ has Lebesgue measure \vspace{1.5mm}zero.
\end{lemma}
\begin{IEEEproof}[\hspace{-1em}Proof]
See Appendix A.
\end{IEEEproof}

In the remainder of our proof that $h (\Pm\ryvb) \rmv >\rmv -\infty$,
%% of Proposition \ref{pro:prelim}, 
we consider an arbitrary $\Qm \!\in\! \sQ$. To use the change-of-variables theorem, we will invoke B\'ezout's theorem to show that the mappings $\phi_{\xv_{\sP}}$ are 
finite-to-one 
\vspace{1.5mm}
a.e.

\begin{lemma}\label{LEMinjective}
Let $\tilde{\sM}$ be defined as the set of all $(\sv,\xv_{\sD})$ such that $\Jm_{\phi_{\xv_{\sP}}}\!(\sv,\xv_{\sD})$ is nonsingular. 
\pagebreak %%%%%%%%
Then for all $\yv\in \phi_{\xv_{\sP}}(\tilde{\sM})$, we 
\vspace{-.5mm}
have 
\be\label{eq:finite}
\abs{\phi_{\xv_{\sP}}^{-1}(\{\yv\})\cap \tilde{\sM}}\,\leq\, \tilde{m}\,\triangleq \,2^{\big(\sum_{t\in[1:\T]}\abs{\sD_t} \,+\, \T\Q\R\big)}\,.
\vspace{1.5mm}
\ee
\end{lemma}

\begin{IEEEproof}[\hspace{-1em}Proof]
Let $\yv \in \phi_{\xv_{\sP}}(\tilde{\sM})$. Then according to \eqref{eq:phi}--\eqref{eq:defXi}, the zeros of the vector-valued mapping 
\[
(\sv,\xv_{\sD})\mapsto \phi_{\xv_{\sP}}(\sv,\xv_{\sD})-\yv
\]
are the common zeros of $\sum_{t\in[1:\T]}\abs{\sD_t}+\T\Q\R$ polynomials of degree 2. Thus, by a weak version of B\'ezout's theorem \cite[Proposition B.2.7]{VdE00}, the number of isolated zeros (i.e., 
%% zeros where there are 
with no other zeros in some neighborhood) cannot exceed $\tilde{m}$. Since $\Jm_{\phi_{\xv_{\sP}}}\!$ 
is nonsingular on $\tilde{\sM}$, the function $\phi_{\xv_{\sP}}$ restricted to $\tilde{\sM}$ is locally one-to-one and, hence, $\phi_{\xv_{\sP}} \!\rmv-\rmv\yv$ has only isolated zeros on $\tilde{\sM}$. Therefore, the number of points $(\sv,\xv_{\sD})\in \tilde{\sM}$ such that $\phi_{\xv_{\sP}}(\sv,\xv_{\sD})=\yv$ cannot exceed $\tilde{m}$. 
%Since these points are exactly those points in the inverse image $\phi_{\xv_{\sP}}^{-1}(\{\yv\})$ that also belong to $\tilde{\sM}$, we conclude that \eqref{eq:finite} holds.
\end{IEEEproof}

Next, we will establish a transformation property of differential entropy under finite-to-one mappings in a general setting. More specifically, we will obtain a lower bound on  differential entropy using the change-of-variables theorem for finite-to-one mappings \cite[Theorem 3.2.5]{fed69} in combination with the uniform bound in 
\vspace{1mm}
Lemma \ref{LEMinjective}.
%%  obtained by B\'ezout's \vspace{1mm}theorem.

\begin{lemma}\label{LEMboundhy}
Let $\ruv\in \IC^n$ be a random vector with continuous density function $f_{\ruv}$. Furthermore, let $\kappa\colon \IC^n \!\rightarrow \IC^n$ be a continuously differentiable mapping with Jacobian matrix $\Jm_{\kappa}$ and let $\sM\triangleq \{\uv \rmv\in\rmv \IC^n \!: \absdet{\Jm_{\kappa}(\uv)}\neq0\}$ and $\rvv \triangleq \kappa(\ruv)$. Assume that the complement of $\sM$ has Lebesgue measure zero and $\abs{\kappa^{-1}(\{\vv\})\cap \sM}\leq m < \infty$ for all $\vv\in \IC^n$, with some constant $m\in \IN$. Then there exists a set $\sU\subseteq \IC^n$ such that 
\vspace{-2mm}
\begin{align*}
h(\rvv) & \,\geq\, - \, m\log (m) - \,m\int_{\sU}f_{\ruv}(\uv)\log (f_{\ruv}(\uv)) \,d\uv\notag \\
& \quad\,\,\ist\ist +\, m\int_{\sU}f_{\ruv}(\uv)\log (\absdet{\Jm_{\kappa}(\uv)}^2) \,d\uv\,. %\label{eq:bounddiffe}
\end{align*}
\vspace{-2mm}
\end{lemma}

\begin{IEEEproof}[\hspace{-1em}Proof]
See Appendix B.
\end{IEEEproof}

To lower-bound $h(\Pm\ryvb\big|\rxv_\sP)$, we first lower-bound the dif\-ferential entropies $h(\Pm\ryvb\big|\rxv_\sP\!\!=\! \xv_\sP)$. By Lemma~\ref{LEMinjective}, we have  $\abs{\phi_{\xv_{\sP}}^{-1}(\{\yv\})\cap \tilde{\sM}}\leq \tilde{m}$. Furthermore, since we assume $\Qm\in \sQ$, we have by Definition \ref{def:qind} that 
$\Jm_{\phi_{\xv_{\sP}}}\!(\sv,\xv_{\sD})$ is nonsingular a.e.\ and, hence, the complement of $\tilde{\sM}$ has Lebesgue measure zero. 
Thus, we can  invoke Lemma~\ref{LEMboundhy} with $h(\rvv)=h(\Pm\ryvb\big|\rxv_\sP\!\!=\! \xv_\sP)$, $\kappa = \phi_{\xv_\sP}$, $\ruv = (\rsv,\rxv_{\sD})$, and $m=\tilde{m}$ to obtain
\vspace{-.5mm}
\begin{multline}
\hspace*{-2mm} h(\Pm\ryvb\big|\rxv_\sP\!= \xv_\sP) \,\geq\, - \,\tilde{m}\log (\tilde{m})\! \\[.5mm]
 \hspace*{-12.5mm} \;\!\rmv- \, \tilde{m}\rmv\int_{\sU}\rmv f_{\rsv,\rxv_{\sD}}(\sv,\xv_{\sD})\log (f_{\rsv,\rxv_{\sD}}(\sv,\xv_{\sD})) \,d(\sv,\xv_{\sD}) \\[1mm]
 \;\; + \,\tilde{m} \rmv\int_{\sU} \rmv f_{\rsv,\rxv_{\sD}}(\sv,\xv_{\sD})\log (\absdet{\Jm_{\phi_{\xv_\sP}}\!(\sv,\xv_{\sD})}^2)\,d(\sv,\xv_{\sD})\,.\label{eq:boundhycondeq}
\vspace{-.5mm}
\end{multline}
%By Lemma \ref{LEMinjective} and Definition \ref{def:qind}, the mapping $\phi_{\xv_\sP}(\sv,\xv_{\sD})$ satisfies the conditions of Lemma \ref{LEMboundhy} for a.a.\ $\xv_\sP$. Hence, 
Using \eqref{eq:boundhycondeq}, we can now lower-bound $h(\Pm\ryvb\big|\rxv_\sP)$ as 
\vspace{-.5mm}
follows:
\begin{multline}
\hspace*{-3mm}h(\Pm\ryvb\big|\rxv_\sP) \,= \int \rmv f_{\rxv_{\sP}}(\xv_\sP) \, h(\Pm\ryvb\big|\rxv_\sP\!= \xv_\sP)\,d\xv_\sP \\ 
\shoveleft\geq \int f_{\rxv_{\sP}}(\xv_\sP) 
\bigg[ \rmv- \ist\tilde{m}\log (\tilde{m}) \\ 
\hspace*{-3.7mm}- \,\tilde{m} \rmv \int_{\sU} \rmv f_{\rsv,\rxv_{\sD}}(\sv,\xv_{\sD})\log (f_{\rsv,\rxv_{\sD}}(\sv,\xv_{\sD})) \,d(\sv,\xv_{\sD}) \\
\;\;\;+ \; \tilde{m} \rmv \int_{\sU} \rmv f_{\rsv,\rxv_{\sD}}(\sv,\xv_{\sD})\log (\absdet{\Jm_{\phi_{\xv_\sP}}\!(\sv,\xv_{\sD})}^2)\,d(\sv,\xv_{\sD}) \bigg]d\xv_\sP \,.\\[-1.5mm]
\label{eq:boundhycond}
\end{multline}
\vspace{-5.5mm}

\noindent
The lower bound in \eqref{eq:boundhycond} consists of three terms. The first term is just a finite constant. 
The second term is finite because the differential entropy of the Gaussian random vector $(\sv,\xv_{\sD})$ is finite. The last term is finite if
\be\label{eq:finlogdet}
\int_{\IC^{\T\L+\T\Q\R}}f_{\rsv,\rxv}(\sv,\xv)\log (\absdet{\Jm_{\phi_{\xv_\sP}}\!(\sv,\xv_{\sD})}^2)\,d(\sv,\xv)
\vspace{.21mm}
\ee
is finite.
To show that \eqref{eq:finlogdet} is finite, we will invoke the following general result for analytic \vspace{1.5mm}functions.

\begin{lemma}\label{LEMboundanalytic}
Let $f$ be an analytic function on $\IC^N$ that is not identically zero. Then
\be\label{eq:expec}
I_1 \ist\triangleq\int_{\IC^N} \!\exp(-\norm{\xiv}^2)\log(\abs{f(\xiv)})\,d\xiv \ist> -\infty \,.
\vspace{1mm}
\ee
\end{lemma}

\begin{IEEEproof}[\hspace{-1em}Proof]
See Appendix C.
\end{IEEEproof}

Since $f_{\rsv,\rxv}$ is the density of a standard multivariate Gaussian random vector and $\det (\Jm_{\phi_{\xv_\sP}}\!(\sv,\xv_{\sD}))$ is a complex polynomial that is not 
identically zero due to the definition of $\sQ$ in Definition \ref{def:qind}, the integral in \eqref{eq:finlogdet} is finite by Lemma \ref{LEMboundanalytic}. Hence, with \eqref{eq:boundhycond}, we obtain $h(\Pm\ryvb\big|\rxv_{\sP})>-\infty$. This concludes the proof that $h (\Pm\ryvb) \rmv >\rmv -\infty$.
%% of Proposition \ref{pro:prelim}.

\vspace{-1mm}

%%%%%%%%%%%%%%%%%%%%%%%%%%%%%%%
\section*{Appendix A:\, Proof of Lemma \ref{LEMnotvanish}} \label{sec:appA}
%%%%%%%%%%%%%%%%%%%%%%%%%%%%%%%

\vspace{.5mm}

We can view $\det(\Jm_{\phi_{\xv_{\sP}}}\!(\sv,\xv_{\sD}))$ as a function $f(\Qm, \xv, \sv)$. Assume that there is a choice of sets $\{\sI_r\}_{r\in[1:\R]}$ satisfying \eqref{eq:indi} and a choice of $\{\sP_t\}_{t\in[1:\T]}$ satisfying \eqref{eq:indp} such that $f(\Qm_0, \xv_0, \sv_0)\neq 0$ at some $(\Qm_0, \xv_0, \sv_0)$. Thus, because for fixed $\xv_0$ and $\sv_0$ the function $f(\Qm, \xv_0, \sv_0)$ is a polynomial in the entries of $\Qm$ and hence analytic in $\Qm$, there is a set $\tilde{\sQ} \!\subseteq\! \IC^{\R\L\times\T\Q}$ 
with a complement of Lebesgue measure zero such that $f(\Qm, \xv_0, \sv_0) \!\neq\! 0$ for all $\Qm \!\in\! \tilde{\sQ}$. 
Hence, for each fixed $\Qm_1 \!\in\! \tilde{\sQ}$, 
%% the function 
$f(\Qm_1, \xv, \sv)$ is not identically zero; furthermore, it is analytic in $\xv$ and $\sv$. 
Therefore, it is nonzero for a.a.\ $(\xv, \sv)$. We conclude that $\Jm_{\phi_{\xv_{\sP}}}\!(\sv,\xv_{\sD})$ is nonsingular and thus $\Qm_1 \!\in\! \sQ$. 
Definition \ref{def:qind} implies that $\tilde{\sQ} \rmv\subseteq\rmv \sQ$, and hence the complement of 
%% the set 
$\sQ$ has Lebesgue measure zero.

It remains to find choices of $\{\sI_r\}_{r\in[1:\R]}$ and $\{\sP_t\}_{t\in[1:\T]}$ such that $f(\Qm, \xv, \sv)\neq 0$ at some $(\Qm, \xv, \sv)$.
We start by choosing sets $\{\sI_r\}_{r\in[1:\R]}$ that satisfy \eqref{eq:indi}. Let 
$k \triangleq \min\big\{\big\lfloor (\T\L-\T)/(\L-\T\Q)\big\rfloor, \R\big\}$
%% \[
%% k\triangleq \min\bigg\{\bigg\lfloor\frac{\T\L-\T}{\L-\T\Q}\bigg\rfloor, \R\bigg\}
%% \]
and $\ell\triangleq \T\L-\T-(\L-\T\Q)\lfloor(\T\L-\T)/(\L-\T\Q)\rfloor$, and 
%% \vspace{-.3mm}
define
%\footnote{Note that the second and third cases in \eqref{eq:indexi} can only occur if $k<\R$; in that case, $k=\lfloor(\T\L-\T)/(\L-\T\Q)\rfloor$ and thus $\ell\geq 0$.}
\be\label{eq:indexi}
\sI_r\triangleq \begin{cases}
[1\!:\!\L], & \text{ if } r \in [1\!:\!k] \\
[1\!:\!\T\Q+\ell], & \text{ if } r=k+1 \\
[1\!:\!\T\Q], & \text{ if } r \in [k+2\!:\!\R].
\end{cases}
\vspace{.3mm}
\ee
For this choice, $[1\!:\!\T\Q]\subseteq\sI_r$ for all $r\in [1\!:\!\R]$, and as many $\sI_r$ as possible without violating \eqref{eq:indi} are equal to $[1\!:\!\L]$. The sets $\{\sP_t\}_{t\in[1:\T]}$ have to satisfy (cf.\ \eqref{eq:indp})
\be
\sum_{t\in[1:\T]}\abs{\sP_t} \,=\, \max\{\T, \T\Q\R-(\R \rmv-\rmv \T)\L\} \,\triangleq\, \vartheta_{\R} \,. 
\label{en:sizep1}
\ee
We define the sets $\sP_t$ 
%% according to the following pattern:
such that $1\in \sP_1$, $2\in \sP_2$, $\dots$, $\T\in \sP_{T}$,
and further 
%% (corresponding to the pairs $(t_i,n_i)$ for $i\in [1\!:\!\T]$) 
$\T \rmv+ 1\in \sP_1$, $\T\rmv+ 2\in \sP_2$, etc., up to $\L\in \sP_{\L \bmod \T}$. 
If 
%% we do not have enough elements to satisfy 
\eqref{en:sizep1} is not yet satisfied, we look for the minimal $t'$ such that $\abs{\sP_{t'}}$ is minimal and $1\!\notin\!\sP_{t'}$ 
and start again with $1 \!\in \sP_{t'}$, $2\in \sP_{t'+1}$, $\dots$ We proceed until \eqref{en:sizep1} is satisfied.
%% we have defined $\max\{\T, \T\Q\R-(\R-\T)\L\}$ elements.
This construction of the sets $\sP_t$ can be formulated 
\vspace*{-1.5mm}
as
%%  (see also Example \ref{EXpt})
\begin{multline}\label{en:defpt}
\sP_t \,\triangleq\, \bigg\{i \rmv\in\rmv  [1\!:\!\L] : \exists\ist j \rmv\in\rmv [1\!\rmv:\!\vartheta_{\R}] \text{ such that }  i\equiv j \bmod \L  \\[-.5mm] 
\text{and } j+\bigg\lfloor \frac{j \!-\! 1}{\lcm(\T,\L)}\bigg\rfloor\!\equiv t \bmod \T\bigg\}\,,
\end{multline}
where $\lcm(\cdot,\cdot)$ denotes the least common multiple.
%%  of $\T$ and $\L$.
For example,
%% \begin{example}\label{EXpt}
for $\T \!=\! \R=3$, $\L \!=\! 6$, and $\Q \!=\! 1$, we have $\vartheta_{\R}=9$ and \eqref{en:defpt} yields
$\sP_1 = \{1, 4, 3 \}$, $\sP_2 = \{2, 5, 1 \}$, and $\sP_3 = \{3, 6, 2 \}$.
%% \begin{align*}
%% \sP_1 & = \{1, \quad 4, \quad \phantom{7,} \quad 3 \} \\
%% \sP_2 & = \{2, \quad 5, \quad 1 \quad\phantom{7,} \}\\
%% \sP_3 & = \{3, \quad 6, \quad 2 \quad\phantom{7,} \}\,.
%% \end{align*}
%% \end{example}
%This definition corresponds to the following procedure. We start with $1\in \sP_1, 2\in \sP_2, \dots, \T\in \sP_{T}$ (corresponding to the pairs $(t_i,n_i)$ for $i\in [1\!:\!\T]$) and proceed with $\T+1\in \sP_1, \T+2\in \sP_2, \dots$ up to $\L\in \sP_{\L \bmod \T}$. If these are not enough elements to satisfy \eqref{en:sizep1} we look for the minimal $t'$ such that $\abs{\sP_{t'}}$ is minimal and $1\notin\sP_{t'}$ and start again with $1\in \sP_{t'}, 2\in \sP_{t'+1}, \dots$ We proceed until we have defined $\max\{\T, \T\Q\R-(\R-\T)\L\}$ elements.
Note that since the sizes of the sets $\sP_t$ differ at most by $1$, \eqref{en:defpt} together with \eqref{en:sizep1} yields 
\begin{align}
\abs{\sP_t} & \,\leq\ist \bigg\lceil \frac{\max\{\T, \T\Q\R-(\R \rmv-\rmv \T)\L\}}{\T} \bigg\rceil \notag\\[.5mm] 
& \,\leq\ist \bigg\lceil \frac{\max\{\T, \T\Q\R-(\R \rmv-\rmv \T)\T\Q\}}{\T} \bigg\rceil \notag \\[.5mm]
& \,=\, \T\Q \,, \label{en:sizept}\\[-6mm]
\notag
\end{align}
where $\L > \T\Q$ has been used.
%% For later use we need some properties of the sets $\sP_t$ for different values of $\R$, which 
Some properties of the sets $\sP_t$ are summarized in the following lemma, whose proof is omitted due to space 
\vspace{1.5mm}
limitations.

\begin{lemma}\label{LEMproppt}
Suppose that $\R \!>\! \T$. Let $\sPt_t \!\in\! [1\!:\!\L]$ be defined according to \eqref{en:defpt} but for $\R \!-\! 1$ receive antennas 
(i.e., $\R$ is formally replaced by $\R \!-\! 1$) and set $\sL_t\triangleq \sPt_t\backslash \sP_t$. Then
\begin{enumerate}[(i)]
\item $\sL_t\cap \sL_{t'}=\emptyset$ for 
\vspace{.5mm}
$t\neq t'$
\item $\sL_t\subseteq \sI_{\R}$
\vspace{.5mm}
\item There exist pairwise disjoint sets $\sG_t$ satisfying 
$\abs{\sG_t}=\Q$, 
$\sG_t\cap \sP_t\neq \emptyset$, and  
$\sG \ist\triangleq\ist \bigcup_{t\in [1:\T]}\sG_t \ist=\ist \sI_{\R}\backslash\bigcup_{t\in [1:\T]}\sL_t$.
\vspace{1.5mm}
\end{enumerate}
\end{lemma}

We will also make repeated use of the following result,
which is a corollary 
\vspace{1.5mm}
of \cite[pp.~21--22]{hojo85}. 

\begin{lemma} \label{LEMdet}
Let $\Mm\in \IC^{N\times N}\rmv$, and let $\sI, \sJ\subseteq [1\!:\!N]$ 
with $|\sI|=|\sJ|$. If ${[\Mm]}_{[1:N]\backslash \sI}^{\sJ} \rmv=\rmv \0v$ or ${[\Mm]}^{[1:N]\backslash \sJ}_{\sI} \!\!=\rmv \0v$, and if ${[\Mm]}_{\sI}^{\sJ}$ is nonsingular, 
then $\det(\Mm) \not= 0$ 
%% does not vanish 
if and only if 
\vspace{1.5mm}
$\det\rmv\Big( {[\Mm]}_{[1:N]\backslash \sI}^{[1:N]\backslash \sJ} \Big) \not= 0$.
%%  does not \vspace{1.5mm}vanish.
\end{lemma}

\begin{remark}
Lemma \ref{LEMdet} is just an abstract way to describe a situation where given a matrix $\Mm$, one is able to make row and column interchanges that yield a new matrix of the form
$
\big(\begin{smallmatrix}
\Am & \Bm \\
\0v & \Cm
\end{smallmatrix}\big)
$
where $\Am$ and $\Cm$ are square matrices.
In this case, it is a basic result that the determinant of $\Mm$ equals the product of the determinants of $\Am$ and $\Cm$.
\end{remark}

%% Recall that we have to choose $\xv$, $\sv$, and $\Qm$ such that $\det(\Jm_{\phi_{\xv_{\sP}}}(\sv,\xv_{\sD}))\neq 0$. 
%% We will conclude the proof of Lemma \ref{LEMnotvanish} 
For the choices of $\{\sP_t\}_{t\in[1:\T]}$ and $\{\sI_r\}_{r\in [1:\R]}$ described above, it now remains to find $\xv$, $\sv$, and $\Qm$ such that 
$f(\Qm, \xv, \sv) = \det(\Jm_{\phi_{\xv_{\sP}}}\!(\sv,\xv_{\sD}))$ is nonzero. This will be done by an induction argument over $\R\geq\T$. 
%Due to space limitations we will only sketch the proof and omit many details.
%Because $\xv$ does not depend on $\R$, we can specify it before we start with the induction argument and set $\xv=(1, \dots, 1)^{\trans}$.

\emph{Induction hypothesis}:
For $\R\geq \T$ (as assumed in Proposition \ref{pro:prelim}), $\{\sP_t\}_{t\in[1:\T]}$ as in \eqref{en:defpt}, and $\{\sI_r\}_{r\in [1:\R]}$ as in \eqref{eq:indexi}, 
there exists a point $(\Qm, \xv, \sv)$ with $\xv=(1, \dots, 1)^{\trans}$ such that $f(\Qm, \xv, \sv)=\det(\Jm_{\phi_{\xv_{\sP}}}\!(\sv,\xv_{\sD}))$ is nonzero.

\emph{Base case (proof for $\R \!=\! \T$)}: 
We have to show that the determinant of the matrix in \eqref{eq:Jacobian3} is nonzero for 
\pagebreak %%%%%%%%%%
$\R \!=\! \T$.
%\be\label{eq:JacobianReqT}
%%\Jm_{\phi_{\xv_{\sP}}}(\sv,\xv_{\sD}) = 
%\left(
%\tilde{\Xim}_1 \,\cdots \, \tilde{\Xim}_{\T} \;
%\begin{matrix}
%\Am_{1,1} & \hspace*{-2mm}\cdots\hspace*{-2mm} & \Am_{1,\T}  \\[-.8mm]
%\vdots    &     & \vdots     \\
%\Am_{\T,1} & \hspace*{-2mm}\cdots\hspace*{-2mm} &\Am_{\T,\T}
%\end{matrix}
%\right),
%\ee
%where (cf.\ \eqref{eq:Jacobian3a}, noting that $\Xm_t=\Iv_{\L}$)
%\[
%\tilde{\Xim}_t\triangleq 
%\begin{pmatrix}
%\Qm_{1,t} \\[-.8mm]
%& \hspace*{-2mm}\ddots\hspace*{-2mm} \\
%&& \Qm_{\T,t}
%\end{pmatrix}.
%\]
For $\R \!=\! \T$, \eqref{en:sizep1} reduces to
$ %\label{eq:sizepcasert}
\sum_{t\in[1:\T]}\abs{\sP_t}= \T^2\Q$,
and with \eqref{en:sizept}, we obtain $\abs{\sP_t}= \T\Q$. Furthermore, from \eqref{eq:indexi}, $\sI_r=[1\!:\!\L]$ for $r\in [1\!:\!\T]$.
We choose $\sv_{r,t}=\0v$ for $r\neq t$, and we choose $[\Qm_{r,t}]_{\sP_r}$ such that 
$
\big[
\big(
\Qm_{r,1}\, \cdots \, \Qm_{r,\T}
\big)
\big]_{\sP_r}
$
is nonsingular.
We have $[\Am_{r,t}]_{\sP_t}=\0v$ (cf.\ \eqref{EQArt}, noting that $\sP_t\cap \sD_t=\emptyset$). Hence, we can use Lemma \ref{LEMdet} with 
$\Mm\triangleq\det(\Jm_{\phi_{\xv_{\sP}}}\!(\sv,\xv_{\sD}))$ 
%% being the matrix in 
given by \eqref{eq:Jacobian3} and 
$
{[\Mm]}_{\sI}^{\sJ} = \diag\big(
\big[
\big(
\Qm_{1,1}\, \cdots \, \Qm_{1,\T}
\big)
\big]_{\sP_1}, \dots, $ \linebreak $
\big[
\big(
\Qm_{\T,1}\, \cdots \, \Qm_{\T,\T}
\big)
\big]_{\sP_\T}
\big)
$.
It thus remains to show 
\vspace*{-.3mm}
that the determinant of the matrix $[\Mm]_{[1:N]\backslash \sI}^{[1:N]\backslash \sJ}$ corresponding 
\vspace*{-.8mm}
to
\be
\label{eq:M_A}
\begin{pmatrix}
{[\Am_{1,1}]}_{\sD_1} \hspace*{-2mm} &\hspace*{-4mm}&\hspace*{-2mm} \\[-1.2mm]
\hspace*{-2mm} & \hspace*{-2mm}\ddots\hspace*{-2mm} \\[0mm]
\hspace*{-2mm} & \hspace*{-4mm} & \hspace*{-2mm} {[\Am_{\T,\T}]}_{\sD_{\T}} 
\end{pmatrix}
\ee
is nonzero. Because of \eqref{EQArt}, this matrix is a diagonal matrix and can be chosen to have nonzero elements by choosing 
${[\Qm_{t,t}]}_{\sD_t}$ and $\sv_{t,t}$ such that $[\Qm_{t,t}]_{\{i\}}\sv_{t,t}\neq 0$ for all $i\in \sD_t$. Thus, the matrix in \eqref{eq:M_A} is
a diagonal matrix with nonzero entries and hence its determinant is nonzero.

\emph{Inductive step}:
We have to show that we can find $\Qm_{\R,t}$ and $\sv_{\R,t}$ for $t \!\in\! [1\!:\!\T]$ such that the determinant of the matrix $\Jm_{\phi_{\xv_{\sP}}}\!(\sv,\xv_{\sD})$ 
in \eqref{eq:Jacobian3} is nonzero assuming that it is nonzero for the $\R \!-\! 1$ setting. 
%Since with increasing $\R$ the sets $\sP_t$ decrease, we define sets $\sPt_t$ which correspond to the definition of the sets $\sP_t$ for $\R-1$. Hence, they are defined by the pairs 
%\[
%(t_i,n_i)=(i+\lfloor i\gcd(\T,\L)/(\T\L)\rfloor \bmod \T, i \bmod \L )
%\]
%by $n_i\in \sPt_{t_i}$ for $i\in [1:\max\{\T, \T\Q(\R-1)-(\R-1-\T)\L\}]$. Since they differ from the sets $\sP_t$ only by additional elements, we have $\sP_t\subseteq\sPt_t$. We define sets $\sL_t\triangleq \sPt_t\backslash \sP_t$. It turns out that by the definition of the sets $\sP_t$ and $\sPt_t$ we have $\sL_t\subseteq \sI_{\R}$ and that the sets $\sL_t$ are pairwise disjoint. Furthermore, the set $\sJ\triangleq \sI_{\R}\backslash\bigcup_{t\in [1:\T]}\sL_t$ can be partitioned into $\T$ sets $\sG_t$ of size $\Q$ satisfying $\sG_t\cap \sP_t\neq \emptyset$. 
Let $\sG$, $\sG_t$, and $\sL_t$ be as in Lemma~\ref{LEMproppt} and let  $g_t \!\in \sG_t \rmv\cap\rmv \sP_t$ ($\not= \emptyset$ due to Lemma~\ref{LEMproppt}). 
Set $[\Qm_{\R,t}]_{\sG\backslash\sG_t}=\0v$. 
Furthermore, let $[\Qm_{\R,t}]_{\sG_t}$ be nonsingular for all $t\in [1\!:\!\T]$.
It easily follows 
%% Thus, we obtain 
that $\big([\Qm_{\R,1}]_{\sG} \,\cdots \, [\Qm_{\R,\T}]_{\sG}\big)$
is nonsingular. Next, we choose $\sv_{\R,t}$ such that it is
%% to be 
orthogonal to the rows of $[\Qm_{\R,t}]_{\sG_t\backslash \{g_t\}}$ and satisfies $[\Qm_{\R,t}]_{\{g_t\}}\sv_{\R,t}\neq 0$. 
With \eqref{EQArt} and $g_t \!\in\! \sP_t$, we then obtain
$[\Am_{\R,t}]_{\sG} \!=\! \0v$, $t \!\in\! [1\!:\!\T]$.
Hence, according to Lemma~\ref{LEMdet} with $\Mm$ given by \eqref{eq:Jacobian3} and 
$[\Mm]_{\sI}^{\sJ}=\big([\Qm_{\R,1}]_{\sG} \,\cdots \, [\Qm_{\R,\T}]_{\sG}\big)$, the determinant of $\Jm_{\phi_{\xv_{\sP}}}\!(\sv,\xv_{\sD})$ in \eqref{eq:Jacobian3} 
is nonzero if and only if the determinant of the following matrix is 
\vspace{.7mm}
nonzero:
\[
%\Jm_{\phi_{\xv_{\sP}}}(\sv,\xv_{\D}) = 
\begin{pmatrix}
 &&& {[\Am_{1,1}]}_{\sI_1} & \hspace*{-2mm}\cdots\hspace*{-2mm} & {[\Am_{1,\T}]}_{\sI_1}  \\[-.5mm]
\hat{\Xim}_1 & \hspace*{-2mm}\cdots\hspace*{-2mm} & \hat{\Xim}_{\T} &  \vdots    &     & \vdots     \\
&& & {[\Am_{\R-1,1}]}_{\sI_{\R-1}} & \hspace*{-2mm}\cdots\hspace*{-2mm} & {[\Am_{\R-1,\T}]}_{\sI_{\R-1}} \\[1.5mm]
\0v & \hspace*{-2mm}\cdots\hspace*{-2mm} & \0v & \hspace*{-2mm}[\Am_{\R,1}]_{\bigcup_{t\in [1:\T]}\sL_t} & \hspace*{-2mm}\cdots\hspace*{-2mm} & [\Am_{\R,\T}]_{\bigcup_{t\in [1:\T]}\sL_t}
\end{pmatrix}\,,
\]
where 
\vspace{1.5mm}
\[
\hat{\Xim}_t \ist\triangleq\ist 
\begin{pmatrix}
\ist{[\Qm_{1,t}]}_{\sI_1} \hspace*{-2mm} &\hspace*{-4mm}&\hspace*{-2mm} \\[-1mm]
\hspace*{-2mm} & \hspace*{-2mm}\ddots\hspace*{-2mm} \\[-.3mm]
\hspace*{-2mm} & \hspace*{-4mm} & \hspace*{-2mm} {[\Qm_{\R-1,t}]}_{\sI_{\R-1}}
\end{pmatrix} .
\]
%% is nonzero. 
By choosing the remaining rows of $\Qm_{\R,t}$ appropriately, we obtain
%% achieve that
$[\Am_{\R,t}]_{\left(\bigcup_{t'\in [1:\T]}\sL_{t'}\right)\backslash \sL_t} \!\rmv=\rmv \0v$
and  $\det\rmv\big({[\Am_{\R,t}]}_{\sL_t}^{\sL_t}\big) \rmv\neq\rmv 0$. By Lemma \ref{LEMdet}, 
%% we finally obtain 
it can then be easily seen that the determinant of $\Jm_{\phi_{\xv_{\sP}}}\!(\sv,\xv_{\sD})$ in \eqref{eq:Jacobian3} is nonzero if and only if
the determinant of \eqref{eq:Jacobian3} for $\R \!-\! 1$ is nonzero, which is true by the induction hypothesis.

\vspace{-3mm}

%%%%%%%%%%%%%%%%%%%%%%%%%%%%%%%
\section*{Appendix B:\, Proof of Lemma \ref{LEMboundhy}} \label{sec:appB}
%%%%%%%%%%%%%%%%%%%%%%%%%%%%%%%

%% \vspace{.5mm}

First, we state the version of the change-of-variables theorem \cite[Theorem 3.2.5]{fed69} that we will \vspace{1.5mm}use.

\begin{lemma}\label{THchange}
Let $\psi\colon \IC^n \!\rightarrow\rmv \IC^n$ be a differentiable mapping with Jacobian matrix $\Jm_{\psi}$. 
Then for any 
\pagebreak %%%%%%%%%
measurable, nonnegative, real-valued function $g$ on $\IC^n$ and any measurable set $\sS\subseteq\IC^n$, we have
\[
\int_{\sS} g(\psi(\uv))\, \absdet{\Jm_{\psi}(\uv)}^2 \, d\uv \,= \int_{\IC^n} \! g(\vv)\operatorname{Nr}(\psi\ist | \ist \sS,\vv) \, d\vv \,,
\]
where $\operatorname{Nr}(\psi\ist | \ist \sS,\vv)$ denotes the number of points $\uv\in \sS$ such that $\psi(\uv)=\vv$. 
(Note, in particular, that $\operatorname{Nr}(\psi\ist | \ist \sS,\vv)=0$ if there is no $\uv\in \sS$ such that $\psi(\uv)=\vv$.)
\end{lemma}

We will also make use of the following lemma to obtain one-to-one mappings with maximal \vspace{1.5mm} support.

\begin{lemma}\label{LEMrestrictinj}
For any Lebesgue measurable set $\sA\subseteq \IC^n$ and any mapping $\psi\colon \IC^n \!\rightarrow\rmv \IC^n$ such that $\abs{\psi^{-1}(\{\vv\})\cap \sA}\leq m < \infty$ for all $\vv\in \IC^n$, there exists a Lebesgue measurable set $\sB \!\subseteq\! \sA$ such that $\psi\big|_\sB$ is one-to-one and $\psi(\sB) \!=\! \psi(\sA)$. 
Furthermore, $\abs{\psi^{-1}(\{\vv\})\cap (\sA\backslash \sB)}\leq m \!-\! 1 < \infty$ for all \vspace{1.5mm}$\vv\in \IC^n$.
\end{lemma}

\begin{IEEEproof}[\hspace{-1em}Proof]
Let $\mathfrak{M}$ denote the set of all measurable subsets $\sV\subseteq \sA$ such that $\psi\big|_\sV$ is one-to-one. We have the natural partial order of inclusion on $\mathfrak{M}$. For any chain (i.e., totally ordered set) $\mathfrak{C}$  of sets in $\mathfrak{M}$, the union of all sets in $\mathfrak{C}$ is an upper bound for all sets in $\mathfrak{C}$ (i.e., for any $\sA_0\in \mathfrak{C}$ we have $\sA_0\subseteq\bigcup_{\sC\in \mathfrak{C}}\sC$) and belongs to $\mathfrak{M}$. Thus, by Zorn's lemma, there exists at least one maximal element in $\mathfrak{M}$. Let $\sB$ be a maximal element in $\mathfrak{M}$. If there exists a $\vv\in \psi(\sA)\backslash \psi(\sB)$, we can add one point $\uv\in \psi^{-1}(\{\vv\})$ to $\sB$ and $\sB\cup \{\uv\}$ belongs to $\mathfrak{M}$ with $\sB\subsetneqq \sB\cup \{\uv\}$, which is a contradiction to the maximality of $\sB$. Hence, $\psi(\sB)=\psi(\sA)$.
Furthermore, since $\sB\in \mathfrak{M}$ the set $\sB$ is measureable and $\psi\big|_\sB$ is one-to-one. Finally,
for each $\vv\in \psi(\sA)$, there exists a $\uv\in \sB$ such that $\psi(\uv) = \vv$. Thus, 
$\abs{\psi^{-1}(\{\vv\})\cap (\sA\backslash \sB)} \leq \abs{\psi^{-1}(\{\vv\})\cap \sA)} - 1 \leq m \!-\! 1$.
\end{IEEEproof}
%\begin{lemma}\label{LEMboundhy}
%Let $\ruv$ be a random variable on $\IC^n$ with continuous density function $f_{\ruv}$. Furthermore, let $\kappa\colon \IC^n \rightarrow \IC^n$ be a continuously differentiable mapping with Jacobian $\Jm_{\kappa}$. Denote with $\sN\triangleq \{\uv\in \IC^n: \det(\Jm_{\kappa})=0\}$ and let $\rvv = \kappa(\ruv)$. Assume that $\sN$ has Lebesgue measure zero and $\abs{\kappa^{-1}(\{\vv\})\backslash \sN}\leq m < \infty$ for all $\vv\in \IC^n$. Then there exists a set $\sU\subseteq \IC^n$ such that 
%%$\tilde{\kappa}\triangleq \kappa\big|_{\sU}$ is one-to-one and $\kappa(\IC^n)\backslash\tilde{\kappa}(\sU)$ has Lebesgue measure zero and such that 
%\be\label{eq:bounddiffe}
%h(\rvv) \geq -m\int_{\sU}f_{\ruv}(\uv)\log f_{\ruv}(\uv) \,d\uv - m\log m+m\int_{\sU}f_{\ruv}(\uv)\log \absdet{\Jm_{\kappa}(\uv)}^2 \,d\uv
%\ee
%holds.
%\end{lemma}

For $\sM$ and $m$ as defined in Lemma~\ref{LEMboundhy}, we now partition the set $\sM$ into subsets $\sV_i$ with $i\in [1\!:\!m]$ such that each $\kappa_i \triangleq\kappa\big|_{\sV_i}$ is one-to-one and $\IC^n\backslash\bigcup_{i\in [1:m]} \sV_i$ has Lebesgue measure zero. The existence of such sets can be shown by using Lemma \ref{LEMrestrictinj} repeatedly.
Next, we define the set $\sU$ used in Lemma~\ref{LEMboundhy}. 
\vspace{-1mm}
Let
\begin{multline}
\tilde{\sU} \,\triangleq\, \bigg\{\uv \rmv\in\rmv \sM: \frac{f_{\ruv}(\uv)}{\absdet{\Jm_{\kappa}(\uv)}^2}\geq \frac{f_{\ruv}(\uvt)}{\absdet{\Jm_{\kappa}(\uvt)}^2} \\
 \forall \,\uvt \in \kappa^{-1}(\kappa(\{\uv\}))\cap \sM\bigg\} \,.\label{eq:Utilde}
\end{multline}

\vspace{-1mm}

\noindent Note that $\kappa(\tilde{\sU})= \kappa(\sM)$. The set $\tilde{\sU}$ is measurable since it is the preimage of $\{1\}$ under the measurable 
\vspace{-.5mm}
function\footnote{The %%%%%%%%%%
function $g$ is measurable by the following argument: $\kappa_i^{-1}$ is continuous by the inverse function theorem. Hence, for all $\uv$ with equal $\sF(\uv)$, the denominator in the definition of $g$ is just the maximum over a finite set of continuous functions and thus measureable. Since there are only a finite number of possible realizations of $\sF(\uv)$, we can partition the domain of $g$ into a finite number of sets where $g$ is measureable. Therefore, $g$ is 
measureable.} %%%%%%%%%
\[
g(\uv) \,\triangleq\, \frac{\frac{f_{\ruv}(\uv)}{\absdet{\Jm_{\kappa}(\uv)}^2}}{\max_{i\in \sF(\uv)}\frac{f_{\ruv}(\kappa_i^{-1}(\kappa(\uv)))}{\absdet{\Jm_{\kappa}(\kappa_i^{-1}(\kappa(\uv)))}^2}} \,,
\]
where $\sF(\uv) \rmv\triangleq\rmv \{i \!\in\! [1\!:\!m] \!: \kappa(\uv) \rmv\in\rmv \kappa_i(\sV_i)\}$. 
By 
\vspace{.3mm}
Lemma~\ref{LEMrestrictinj} %%\linebreak %%%%%%%%% 
with $\psi \!=\! \kappa$ and $\sA \!=\rmv \tilde{\sU}$, there exists a 
\pagebreak %%%%%%%%
set $\sU \rmv\subseteq \tilde{\sU}$ such that $\kappa\big|_{\sU}$ is one-to-one and 
$\kappa(\sU) \rmv=\rmv \kappa(\tilde{\sU}) \rmv=\rmv \kappa(\sM)$.
Applying Lemma~\ref{THchange}\linebreak %%%%%%%%%  
with $g(\vv) \rmv=\rmv - f_{\rvv}(\vv)\log (f_{\rvv}(\vv))$, $\psi \rmv=\rmv \kappa$, and $\sS \rmv\rmv=\rmv \sU$ yields\linebreak %%%%%%%%% 
($f_{\rvv}$ denotes the density of $\rvv=\kappa(\ruv)$)
\begin{align}
h(\rvv) & \,=\, -\int_{\IC^n} \! f_{\rvv}(\vv)\log (f_{\rvv}(\vv))\, d\vv \notag \\[1mm]
& \stackrel{(a)}{=} -\int_{\IC^n} \! f_{\rvv}(\vv)\log (f_{\rvv}(\vv)) \operatorname{Nr}(\kappa\ist | \ist \sU,\vv) \,d\vv \notag \\[1mm]
& \,=\, -\int_\sU f_{\rvv}(\kappa(\uv))\log (f_{\rvv}(\kappa(\uv))) \,\absdet{\Jm_{\kappa}(\uv)}^2 \, d\uv \,. \label{eq:hpmy}
\end{align}
Here, $(a)$ holds because $\vv$ is supported (up to a set of measure zero) on $\kappa(\sU)$; note also 
that $\operatorname{Nr}(\kappa\ist | \ist \sU,\vv)$ is $1$ for $\vv = \kappa(\uv)$ and $0$ else.
The next step is to establish a relation between the densities $f_{\rvv}(\kappa(\uv))$ and $f_{\ruv}(\uv)$ for $\uv\in \sU$. 
Let $\sU' \!\rmv\subseteq\rmv \sU$ be any measurable subset of $\sU$. We have 
\begin{align}
\int_{\sU'} \! f_{\rvv}(\kappa(\uv))\,\absdet{\Jm_{\kappa}(\uv)}^2\, d\uv 
& \,= \int_{\kappa(\sU')} \!f_{\rvv}(\vv)\, d\vv \notag\\[.5mm]
& \,=\, \operatorname{Pr}\{\vv\in \kappa(\sU')\} \notag\\[.5mm]
& \,=\, \operatorname{Pr}\{\uv\in \kappa^{-1}(\kappa(\sU'))\} \notag\\[.5mm]
& \,= \int_{\kappa^{-1}(\kappa(\sU'))}\!f_{\ruv}(\uv)\, d\uv \,. \label{eq:dens1}\\[-7mm]
%& \leq \sum_{i\in [1:m]}\int_{\kappa_i^{-1}(\kappa(\sU'))}\!\!f_{\ruv}(\uv)\, d\uv. \label{eq:dens1}
\notag
\end{align}
Since $\kappa_i=\kappa\big|_{\sV_i}$, we have 
\[
\bigcup_{i\in [1:m]} \! \kappa_i^{-1}(\kappa(\sU')) \,=\rmv \bigcup_{i\in [1:m]} \! (\kappa^{-1}(\kappa(\sU'))\cap \sV_i) \,,
\vspace{.5mm}
\]
and since $\IC^n\backslash\bigcup_{i\in [1:m]} \sV_i$ has Lebesgue measure zero, the set $\bigcup_{i\in [1:m]}\kappa_i^{-1}(\kappa(\sU'))$ 
is equal to $\kappa^{-1}(\kappa(\sU'))$ up to a set of Lebesgue measure zero. Thus, 
\be
\int_{\kappa^{-1}(\kappa(\sU'))} \! f_{\ruv}(\uv)\, d\uv \,=\rmv \sum_{i\in [1:m]}\int_{\kappa_i^{-1}(\kappa(\sU'))} \! f_{\ruv}(\uv)\, d\uv  %\label{eq:dens1}
\vspace{-.5mm}
\ee
(note that $\kappa_i^{-1}(\kappa(\sU')) \subseteq \sV_i$ and the $\sV_i$ are disjoint).
Using for an arbitrary $i\in [1\!:\!m]$ Lemma \ref{THchange} with $\psi = \kappa_i^{-1}$ and  $\sS=\kappa_i(\kappa_i^{-1}(\kappa(\sU')))$, and using the inverse function theorem,
%with the functions $\psi = \kappa_i^{-1}\circ\tilde{\kappa}$ and $g=f_{\ruv}$ and the sets $\sA=\tilde{\kappa}^{-1}(\kappa_i(\kappa_i^{-1}(\kappa(\sU'))))$ 
we obtain (note that $\Jm_{\kappa} \rmv=\rmv \Jm_{\kappa_i}$ on $\sV_i$ because 
\vspace{-1mm}
$\kappa_i=\kappa\big|_{\sV_i}$)
\be\label{eq:dens2a}
\int_{\kappa_i^{-1}(\kappa(\sU'))}\!f_{\ruv}(\uv)\, d\uv
 \,= \int_{\kappa_i(\kappa_i^{-1}(\kappa(\sU')))}\! \frac{f_{\ruv}(\kappa_i^{-1}(\vv))}{\absdet{\Jm_{\kappa}(\kappa_i^{-1}(\vv))}^2}\, d\vv \,.
\vspace{.5mm}
\ee
Another application of Lemma \ref{THchange} with $\psi = \tilde{\kappa} \triangleq \kappa\big|_\sU$ and $\sS = \tilde{\kappa}^{-1}(\kappa_i(\kappa_i^{-1}(\kappa(\sU'))))$ then gives
\begin{align}
&\hspace{-2mm}\int_{\kappa_i(\kappa_i^{-1}(\kappa(\sU')))} \frac{f_{\ruv}(\kappa_i^{-1}(\vv))}{\absdet{\Jm_{\kappa}(\kappa_i^{-1}(\vv))}^2}\, d\vv \notag \\[.6mm]
& \hspace{1mm}= \int_{\tilde{\kappa}^{-1}(\kappa_i(\kappa_i^{-1}(\kappa(\sU'))))} \!\!\frac{f_{\ruv}(\kappa_i^{-1}(\tilde{\kappa}(\uvt)))\,\absdet{\Jm_{\kappa}(\uvt)}^2}{\absdet{\Jm_{\kappa}(\kappa_i^{-1}(\tilde{\kappa}(\uvt)))}^2}\, d\uvt \,. \label{eq:dens2}
\end{align}
We can upper-bound \eqref{eq:dens2} by
\begin{align}
& \int_{\tilde{\kappa}^{-1}(\kappa_i(\kappa_i^{-1}(\kappa(\sU'))))} \!\!\frac{f_{\ruv}(\kappa_i^{-1}(\tilde{\kappa}(\uvt)))\,\absdet{\Jm_{\kappa}(\uvt)}^2}{\absdet{\Jm_{\kappa}(\kappa_i^{-1}(\tilde{\kappa}(\uvt)))}^2}\, d\uvt \notag \\[.8mm]
& \quad \stackrel{(a)}\leq\! \int_{\tilde{\kappa}^{-1}(\kappa_i(\kappa_i^{-1}(\kappa(\sU'))))} \!\!\frac{f_{\ruv}(\uvt)\,\absdet{\Jm_{\kappa}(\uvt)}^2}{\absdet{\Jm_{\kappa}(\uvt)}^2}\, d\uvt \notag \\[.2mm]
& \quad \ist\ist= \int_{\tilde{\kappa}^{-1}(\kappa_i(\kappa_i^{-1}(\kappa(\sU'))))} \!f_{\ruv}(\uvt) \, d\uvt \notag\\[.5mm]
& \quad \ist\stackrel{(b)}\leq\! \int_{\sU'} \! f_{\ruv}(\uvt) \, d\uvt \,, \label{eq:dens3}
\end{align}
where in $(a)$ we used the fact that $\uvt\in\tilde{\sU}$ (we have 
$\uvt \in \tilde{\kappa}^{-1}(\kappa_i(\kappa_i^{-1}(\kappa(\sU')))) 
= \tilde{\kappa}^{-1}(\kappa_i(\kappa_i^{-1}(\tilde{\kappa}(\sU'))))
= (\tilde{\kappa}^{-1}\circ\kappa_i)((\tilde{\kappa}^{-1}\circ\kappa_i)^{-1}(\sU'))
\subseteq \sU' \!\subseteq\ist \sU\subseteq\ist \tilde{\sU}$) and the inequality in \eqref{eq:Utilde}, and in $(b)$ we used $\tilde{\kappa}^{-1}(\kappa_i(\kappa_i^{-1}(\kappa(\sU'))))\subseteq\sU'$. Note that the upper bound \eqref{eq:dens3} does not depend on $i\in [1\!:\!m]$. Hence, \eqref{eq:dens1}--\eqref{eq:dens3} yield
\[
%% \label{eq:dens}
\int_{\sU'} \! f_{\rvv}(\kappa(\uv))\,\absdet{\Jm_{\kappa}(\uv)}^2 \, d\uv \,\leq\, m \rmv \int_{\sU'} \! f_{\ruv}(\uv) \, d\uv\,,
\]
for an arbitrary measurable set $\sU' \!\subseteq \sU$. Thus,
\[
 f_{\rvv}(\kappa(\uv))\,\absdet{\Jm_{\kappa}(\uv)}^2 \ist\leq\, m\, f_{\ruv}(\uv) \quad \text{a.e.\ on } \sU \,.
\]
Inserting this into \eqref{eq:hpmy} leads to
\begin{align*}
h(\rvv) & \,\geq\, -\int_\sU \rmv m \,f_{\ruv}(\uv) \log \rmv\bigg(\frac{m f_{\ruv}(\uv)}{\absdet{\Jm_{\kappa}(\uv)}^2}\bigg) \, d\uv \\[1mm]
 & \,\geq\,  -~m\log (m) - m \rmv \int_{\sU} \rmv f_{\ruv}(\uv)\log (f_{\ruv}(\uv)) \,d\uv \\
 & \quad\;\;\rmv +\ist m \rmv \int_{\sU} \rmv f_{\ruv}(\uv)\log (\absdet{\Jm_{\kappa}(\uv)}^2) \,d\uv \,.\\[-3mm]
\end{align*}

\vspace{-5mm}

%%%%%%%%%%%%%%%%%%%%%%%%%%%%%%%
\section*{Appendix C:\, Proof of Lemma \ref{LEMboundanalytic}} \label{sec:appC}
%%%%%%%%%%%%%%%%%%%%%%%%%%%%%%%

\vspace{1mm}

Since $f$ is not identically zero, there is a $\xiv_0 \!\in\rmv \IC^{N}\rmv$ such that $f(\xiv_0)\neq 0$. 
%% Suppose that $f(\xiv_0)\neq 0$ for some $\xiv_0\in \IC^{N}$.
Then $g(\xiv)\triangleq f(\xiv+\xiv_0)$ is an analytic function that is nonzero at $\xiv=\0v$.
By changing variables $\xiv \mapsto \xiv +\xiv_0$, we obtain for $I_1$ in \eqref{eq:expec}
\[
%% \label{eq:bound1a}
I_1 \ist=\int_{\IC^N} \! \exp(-\norm{\xiv+\xiv_0}^2)\log(\abs{g(\xiv)})\,d\xiv\,.
\vspace{-1mm}
\]
Noting that
\begin{align*}
\norm{\xiv+\xiv_0}^2 & \ist\leq\, \norm{\xiv}^2+2\norm{\xiv}\norm{\xiv_0}+\norm{\xiv_0}^2 \\
& \ist\leq\, \norm{\xiv}^2+2\max \{\norm{\xiv}^2, \norm{\xiv_0}^2\}+\norm{\xiv_0}^2 \\
& \ist\leq\, 3\norm{\xiv}^2+ 3\norm{\xiv_0}^2\,,
\end{align*}
we can lower bound $I_1$ by 
\be\label{eq:bound1}
I_1 \ist\geq\,
c \rmv \int_{\IC^N} \!\exp(-3\norm{\xiv}^2)\log(\abs{g(\xiv)})\,d\xiv \,\triangleq\, I_2\,,
\vspace{-.5mm}
\ee
with $c\triangleq \exp(-3\norm{\xiv_0}^2)$.
%By \cite[Example 2.6.1.3]{Azarin09} the function $\log(\abs{g(\xiv)})$ is subharmonic 
%as a function of all $2N$ real and imaginary parts of $\xiv$. 
Using
%% With 
the mapping $\varphi \colon \IR^{2N} \!\!\rightarrow\rmv \IC^N$;
$\xv \mapsto \big(\xv_{[1:N]}+i\xv_{[N+1:2N]}\big)$,
%% \[
%% \varphi\colon \begin{cases}
%% \IR^{2N} &\hspace*{-3mm}\rightarrow \IC^N \\
%% \xv &\hspace*{-3mm} \mapsto \big(\xv_{[1:N]}+i\xv_{[N+1:2N]}\big)\,,
%% \end{cases}
%% \]
we can write $I_2$ in \eqref{eq:bound1} as
\be\label{eq:subharm}
I_2 \,=\, c \rmv \int_{\IR^{2N}} \!\exp(-3\norm{\xv}^2)\,u(\xv)\,d\xv\,,
\ee
with $u(\xv)\triangleq \log(\abs{g(\varphi(\xv))})$. Since $g(\0v) \!\neq\! 0$, we have $u(\0v)>-\infty$.
By \cite[Example 2.6.1.3]{Azarin09}, $u(\xv)$ is a subharmonic function. 
A useful property of subharmonic functions is stated in the following lemma 
\vspace{1.5mm}
(see \cite[Theorem 2.6.2.1]{Azarin09}).

\begin{lemma}\label{lem:subharmonic}
Let $u$ be a subharmonic function on $\sW\subseteq \IR^{2N}\rmv$, and let $\xv \!\in\! \IR^{2N}\rmv$.
If $\sB_{\xv, r} \!\subseteq\! \sW$ for some $r \!>\! 0$, with $\sB_{\xv, r}\triangleq\{\vv \rmv\in\rmv \IR^{2N} \!: \norm{\vv \rmv-\rmv \xv}\leq r\}$, 
\pagebreak %%%%%%%
then 
\[
u(\xv) \,\leq\, \frac{1}{\sigma_{2N}\, r^{2N-1}} \rmv \int_{\sS_{\xv,r}}\!\rmv u(\yv) \,ds(\yv)\,,
\]
where $\sS_{\xv,r}\triangleq \{\yv \rmv\in\rmv \IR^{2N} \!: \norm{\yv \rmv-\rmv \xv} \rmv=\rmv r\}$, $\sigma_{2N}$ is the area of the unit sphere in $\IR^{2N}\rmv$, 
and $ds$ denotes integration with respect to the $(2N \!-\! 1)$-dimensional Hausdorff measure (cf.\ \cite[Subsection 2.10.2]{fed69}).
\vspace{1.5mm}
\end{lemma}

Using a well-known measure-theoretic result \cite[Theorem~3.2.12]{fed69}, we obtain
\begin{multline}\label{eq:help26}
\int_{\IR^{2N}} \!\exp(-3\norm{\xv}^2) \,u(\xv)\,d\xv \\
 =
\int_{(0,\infty)} \int_{\sS_{\0v,r}} \!\rmv\exp(-3 r^2)\, u(\yv)\,ds(\yv)\,dr \,.
\end{multline}

\vspace*{-1mm}

\noindent
We thus 
\vspace*{.5mm}
have
\begin{align*}
I_2 
& \,\stackrel{(a)}=\, c \rmv \int_{(0,\infty)} \int_{\sS_{\0v,r}} \!\rmv\exp(-3 r^2) \,u(\yv)\,ds(\yv)\,dr \\[.3mm]
& \,\stackrel{(b)}\geq\, c\, \sigma_{2N}\, u(\0v)\int_{(0,\infty)}\!\rmv \exp(-3 r^2) \, r^{2N-1} \ist dr \\[-.3mm]
& \,\stackrel{(c)}>\, -\infty\,,
\end{align*}
where $(a)$ follows by using \eqref{eq:help26} in \eqref{eq:subharm}, 
$(b)$ is due to Lemma \ref{lem:subharmonic}, and $(c)$ holds because $u(\0v) \!>\! -\infty$. With \eqref{eq:bound1}, it then follows that \vspace{1mm} $I_1>-\infty$.

\vspace{-1.5mm}

%%%%%%%%%%%%%%%%%%%%%%%%%%%%%%%
\section*{Acknowledgment} 
%%%%%%%%%%%%%%%%%%%%%%%%%%%%%%%

\vspace{.5mm}

We wish to thank Dr. Shaowei Lin for kindly pointing us to the weak version of 
B\'ezout's theorem.

%% \vspace{1mm}

\renewcommand{\baselinestretch}{1.07}\small\normalsize

\bibliography{references}
\bibliographystyle{IEEEtran}
\end{document}